\documentclass[a4paper,12pt]{article}
\usepackage{latexsym,amsmath,tabularx,amssymb,bm}
\usepackage{cite}

\newcommand{\tr}{{\rm tr}}               

\newcommand{\nn}{\nonumber\\ }
\newcommand{\be}{\begin{eqnarray}}
\newcommand{\ee}{\end{eqnarray}}

\def\bfomega{\mbox{\boldmath$\omega$}}
\def\bftheta{\mbox{\boldmath$\theta$}}

\def\simge{\mathrel{%
    \rlap{\raise 0.511ex \hbox{$>$}}{\lower 0.511ex \hbox{$\sim$}}}}
\def\simle{\mathrel{
    \rlap{\raise 0.511ex \hbox{$<$}}{\lower 0.511ex \hbox{$\sim$}}}}
\def\bigs{\mathrel{
    \rlap{\raise 0.531ex \hbox{$>$}}{\lower 0.531ex \hbox{$<$}}}}

\def\del{\partial}                              

\def\frac#1#2{{#1 \over #2}}

\def\half{\ifinner {\scriptstyle {1 \over 2}}
    \else {1 \over 2} \fi}

\newif\ifpdf
\ifx\pdfoutput\undefined
      \pdffalse                           
      
      \usepackage{graphicx}
\else
      \pdfoutput=1
      \pdftrue
      \usepackage[pdftex]{graphicx}
      
      \usepackage[backref,colorlinks=true]{hyperref}
\fi

\begin{document}
\begin{flushright}
\hfill Saclay-T01/136\\
\hfill NSF-ITP-02-49
\end{flushright}
\vspace{.6cm}

\noindent{\Large\bf Non linear gluon evolution in path-integral form}

   \vspace{0.5cm}
   {\large Jean-Paul Blaizot$^{\rm a}$,
 Edmond Iancu$^{\rm a,c}$ and Heribert Weigert$^{\rm b}$}

\vspace{0.1cm}
{\it $^{\rm a}$~Service de Physique Th{\'e}orique, CEA/DSM/SPhT,
Unit{\'e} de recherche associ{\'e}e au CNRS, CEA-Saclay, F-91191 
        Gif-sur-Yvette cedex, France}

\vspace{0.1cm}

{\it $^{\rm b}$~Universit{\"a}t Regensburg, 93040 Regensburg, 
Germany  }

\vspace{0.1cm}

{\it $^{\rm c}$~Institute for Theoretical Physics, 
                  University of California, 
                  Santa Barbara,\\ CA 93106-4030, USA 
        }

\vspace{.2cm}
\noindent\begin{center}
\begin{minipage}{.91\textwidth}
  {\small{\bf Abstract.}
We explore and clarify the connections between two 
different forms of the renormalisation group
equations describing the quantum evolution of 
hadronic structure functions at small $x$. This 
connection is established via a Langevin 
formulation and associated path integral solutions that
highlight  the statistical nature of the quantum evolution, 
pictured here as a random walk in the space of Wilson lines. 
The results confirm known approximations, form the basis for 
numerical simulations and widen
the scope for further analytical studies. }
\end{minipage}
\end{center}
\setcounter{equation}{0}

\section{Introduction}

A lot of effort is presently devoted to the study of hadronic systems
with large parton densities, as revealed for instance in high energy
scattering experiments. Of particular interest is the regime of {\em
   parton saturation} \cite{GLR,MQ,BM87,AM0,MV94,JKMW97,AM2,SAT,GBW99}
expected to end the growth with increasing energy of the structure
functions  which is predicted by linear
evolution equations \cite{BFKL,DGLAP} and is experimentally
observed at HERA.  These linear equations are in fact only justified in
situations where the partons in the hadrons
form a dilute system; they single out two-particle
correlators (actually the single particle densities) as the only important
contributions. In the high density regime however, all $n$-point functions
have to be treated on the same footing, and  the ensuing evolution is
  non-linear.
  Several formalisms have been developed recently to cope with this non
linear regime
\cite{MV94,JKMW97,K96,KM98,AM3,JKLW97,JKW99,B,K,W,Braun,KMW00,PI,AM01,BB01}.
(See also Refs. \cite{CARGESE,AMCARGESE} for recent reviews and more
references.)

To be specific we shall focus on deep inelastic scattering where a
virtual photon collides with a hadronic target. The crucial variable
in the discussion is the large rapidity separation $\tau\equiv \ln(1/x)\sim
\ln s$ between the projectile and the target; $s$~is the total energy
squared and $x$ is the typical fraction of the hadron momentum carried by the partons seen by the
photon. Large rapidity interval implies ``small $x$''and all formalisms exploit this kinematic
situation to reduce the number, and select the type, of degrees of freedom involved in the event.

A key simplification comes from the recognition that the basic degrees of
freedom, in an appropriate frame and for a suitable gauge, involve only
the   component $A^+(x^-,{\bm{x}})$ of the gauge field of
the hadron, and
this enters only through path ordered exponentials along $x^-$ (Wilson
lines):
\begin{equation}\label{Udagger}
U^\dagger({\bm{x}})\,\equiv\,{\rm P}\,{\rm exp}\left({\rm i}g\int dx^-
A^+_a(x^-,{\bm{x}}) t^a\right). 
\end{equation}
(Throughout, we shall use light-cone coordinates, e.g.,
$k^\mu=(k^+,k^-,{\bm{k}})$, with $k^\pm\equiv (k^0\pm k^3)/\sqrt 2$, and
${\bm{k}}$ is a vector in a plane perpendicular to the direction of
propagation of the photon, to which we shall refer briefly as the
``transverse plane".) Thus the total cross section in deep inelastic
$e A$ ($\gamma^* A$) scattering is of the schematic form
($\bm{r}=\bm{x}-\bm{y}$, $\bm{b}=\bm{x}+\bm{y}$) \cite{AM0,NZ91,GBW99}:
\begin{equation}
   \label{eq:sigma_tot}
   \sigma_{\gamma^* A}(x,Q^2)=
2\int\!\!dz \int\!\!d^2{\bm{r}}\ \vert \psi(z,{\bm
r};Q^2)\vert^2 \int\!\!d^2{\bm b}\
    (1-S_\tau({\bm{x}},{\bm{y}}))
\end{equation}
which naturally
separates into a  photon wave function $\psi$, giving the probability that
the virtual photon splits into  a $q\bar q$ pair (a color ``dipole'')
with transverse size ${\bm{r}}$, and a matrix element summarizing the QCD
interaction of the dipole with the hadronic target:
\be\label{Stau}
S_\tau({\bm{x}},{\bm{y}})\,\equiv\,\frac{1}{N_c}\,
\langle \tr\big(U^\dagger({\bm{x}}) U({\bm{y}})\big)
\rangle_\tau.\ee
This form ensues if the energy is large enough for the scattering to
be treated in the eikonal approximation with the $q\bar q$ pair not
deflected from its light-like trajectory.

At this point, although one has achieved a reduction of degrees of
freedom, the calculation of the average $\langle \ldots\rangle$ in
Eq.~(\ref{Stau}) is still a very hard problem. Note that this average
depends on the rapidity interval $\tau$, and   rather than calculating directly the average,
one can at least try to calculate its variation with $\tau$. This
calculation  has been approached by following two seemingly distinct
routes. In the  first one, one focuses on the averages of products of
Wilson lines and obtains differential
equations for their evolution with $\tau$
without having to  specify explicitly the average on the target hadron
\cite{B,K,AMCARGESE}.  Such equations form an
infinite hierarchy coupling together correlators
of Wilson lines of increasing order
\cite{B} (see also Refs. \cite{K,Braun,BB01,AMCARGESE}).
Remarkably, this hierarchy  is coded in a single
functional evolution equation for an appropriate weight function
\cite{W} (see also Ref. \cite{AM01}). 
The second approach builds on an explicit model for the
hadron wavefunction which allows for a direct and explicit
calculation of the average
\cite{MV94,JKMW97,K96,KM98,JKLW97,JKW99,KMW00,PI,SAT}.

In both approaches, the evolution results from a renormalization group (RG)
procedure in which  quantum fluctuations of the gluonic fields are
successively integrated out \cite{JKLW97,KMW00,PI}.
Reassuringly, as shown in Ref.~\cite{PI}, both approaches yield identical
results for the calculation of observables. Both lead to a description of
quantum evolution of parton densities in terms of  equations for
probability distributions. A complete derivation and thorough proof of
these correspondences, however, has not been available up to now; this
is the main purpose of the current publication.

As we shall see, both pictures of the evolution can be vizualized as a
random walk in the space of Wilson lines. For each value of $\tau$, there
exists some probability that a particular value of $U$ is realized,
i.e., there is an associated probability distribution $Z_\tau[U]$ that allows the calculation of
averages such as that in Eq.~(\ref{Stau}). This distribution satisfies a Fokker-Planck equation: this
is the evolution equation of the first approach, which we shall refer too as the
``$U$-representation''. If we go deeper in the details of the random
walk, one finds that the value of $U$ at rapidity $\tau+{\rm d}\tau$, call
it $U_{\tau+{\rm d}\tau}$, is related to $U_{\tau}$ by $U_{\tau+{\rm
    d}\tau}=U_{\tau}\exp(-i\alpha_\tau{\rm d}\tau) $,
 where $\alpha_\tau$ is a random variable --- the component $A^+$ of a classical
colour field ---
whose probability distribution $W_\tau[\alpha]$ is explicitly calculated in
the second approach which provides also the corresponding evolution
equation.  We shall refer to this second approach as the
``$\alpha$-representation''.  The correspondence between the two approaches is
easy to see from the point of view of the $\alpha$-representation. Indeed,
in this representation, the matrix $U$ at rapidity $\tau$ is constructed
step by step from some initial $U_0$ by (right)
multiplication by a matrix of
the form $\exp(-i\alpha_\eta)$ when
$\alpha_\eta$ is a random variable. Since we know
the probability distribution of all the random variables $\alpha_\eta$ where
$\eta$ spans the rapidity interval --- this is $W_\tau[\alpha]$ --- one can
easily calculate the probability $Z_\tau[U]$ that the random walk
terminates at $U$. Furthermore, one can relate the two evolution
equations.

Further insight is gained by considering the random steps as a
Langevin process, and by writing the average over the random variables
as path integrals.  Such path integrals provide a
direct connection between the two approaches.  This not only
highlights the physical concepts driving the RG procedure with a clear
reference to its statistical nature, but it also allows us to give
simple formulae for the Langevin process, which may be convenient for
numerical simulations.  Investigations along these lines have actually
already been carried out and will be available soon~\cite{HWR}.
In Ref. \cite{Balitsky2001}, a different path integral has been proposed
as a solution to the infinite hierarchy of coupled equations established
in \cite{B}. The precise relation between the results in \cite{Balitsky2001}
and the results that we shall obtain below
in this paper is still to be explored.

In the next section we shall elaborate on the physics underlying
the RG equations of Refs~\cite{W} and~\cite{PI} and recall the explicit
forms of the basic equations in the two representaions introduced
above. The formal relation between the two formalisms will be
outlined. In particular the characteristics of the random process
responsible for the evolution with rapidity will put forward.  Then,
in section \ref{sec:part-brownian} we show that much of the structure
of the evolution equations can be identified in the simpler setting of
the random walk in ordinary Euclidean space. The last section
considers finally the the relevant case of random walks on a group
manifold. Conclusions are summarized at the end.

\setcounter{equation}{0}
\section{Nonlinear quantum evolution in two representations}
\label{sec:RGE}

As discussed in the previous section, the calculation of
scattering observables at high energy reduces to that of averages of
products of Wilson lines. This can be implemented in several
ways, and we shall review in this section two seemingly independent
approaches which turn out to be equivalent. Our goal here is to recall
the main physical ingredients. The rest of the paper will be devoted to a
deeper understanding of the relation between the two approaches by
analyzing their common mathematical grounds.

\subsection{Wilson line operators and the $U$--representation}

Consider the evolution of the cross-section  (\ref{eq:sigma_tot})
when the rapidity gap $\tau$  increases by an amount $d\tau$.
We choose the frame so that
$\tau_H\approx \tau \gg  \tau_{dipole}$, where $\tau_H$ and
$\tau_{dipole}$ are the rapidities of the hadron and the dipole,
respectively.  If initially
$\alpha_s\tau_{dipole}\ll 1$, the dipole is just a quark-antiquark pair
\cite{AMCARGESE}. When $\tau_{dipole}$ is increased by an
amount $d\tau$, so is the phase space for the emission of a gluon. If
such a gluon is indeed emitted, then the interaction of the dipole with
the target will involve the independent scattering of the quark, the
antiquark, {\em and} the emitted gluon, off the
colour field
$A(x^-, {\bm{x}})$ of the target.  This is the source for the change 
$dS_\tau\propto
\alpha_s d\tau$ in the $S$--matrix element
$S_\tau\equiv\frac{1}{N_c}\langle{\rm tr}(U^\dagger_{\bm{x}} U_{\bm{y}})
\rangle_\tau$, for which the following evolution equation is
obtained:
\begin{align}\label{evolV}
\partial_\tau \langle {\rm tr}(U^\dagger_{\bm{x}} U_{\bm{y}})
\rangle_\tau=&
\nonumber \\
-{\alpha_s\over 2 \pi^2}\int\! d^2{\bm{z}}
& \,\frac{(\bm{x}-\bm{y})^2}{(\bm{x}-\bm{z})^2(\bm{y}-\bm{z})^2 
}
\left\langle N_c {\rm tr}(U^\dagger_{\bm{x}} U_{\bm{y}})
- {\rm tr}(U^\dagger_{\bm{x}} U_{\bm{z}})
{\rm tr}(U^\dagger_{\bm{z}} U_{\bm{y}})\right\rangle_\tau,
\end{align}
where all the Wilson lines are in the fundamental representation, and
$U_{\bm{x}}\equiv U({\bm{x}})$. Eq.~(\ref{evolV}) has been originally 
derived by Balitsky
\cite{B}.

While  this
equation for $S_\tau(\bm{x},\bm{y})$ involves only
averages of products of Wilson lines, this is not a closed equation:
the 2-point function is related to a 4-point function. However a 
closed equation is
obtained when one allows $N_c$ to take arbitrary large values. 
Indeed, in this large $N_c$
limit,  the product of traces in its r.h.s. factorizes,
\be
\left\langle {\rm tr}(U^\dagger_{\bm{x}} U_{\bm{z}})
{\rm tr}(U^\dagger_{\bm{z}} U_{\bm{y}})\right\rangle_\tau
\longrightarrow
\left\langle {\rm tr}(U^\dagger_{\bm{x}} U_{\bm{z}})\right\rangle_\tau\,
\left\langle {\rm tr}(U^\dagger_{\bm{z}} U_{\bm{y}})
\right\rangle_\tau\quad {\rm for}\,\,N_c\to\infty,
\ee
so that eq.~(\ref{evolV}) reduces indeed to a closed equation for 
$S_\tau(\bm{x},\bm{y})$.
In this form, the equation has been independently obtained by
Kovchegov\cite{K},
within the Mueller's dipole model \cite{AM3}. It
has attracted much interest recently \cite{LT99,AB01,GB01,SCALING}.(See 
also Ref. \cite{Braun}
for still a different derivation,
which is summing ``fan'' diagrams.)

In general, eq.~(\ref{evolV}) is just the first step
in an infinite hierarchy of coupled evolution
equations for the correlation functions
of Wilson lines. In principle, these equations can be all obtained
in the formalism of Ref. \cite{B}, but, as recognized in Ref. \cite{W},
they are equivalent
to a single evolution equation for an appropriate
generating functional $Z_\tau[U]$ giving the probability that a 
certain field
$U_{\bm{x}}
$ is realized in a collision with rapidity gap $\tau$. In particular 
the averages of
products of Wilson lines are obtained as:
\begin{equation}
   \label{eq:expect-val}
   \langle U_{\bm{x}_1}^{(\dagger)} \ldots U_{\bm{x}_n}^{(\dagger)}
\rangle_\tau
   = \int \!\![{\rm d}\mu(U)]\ U_{\bm{x}_1}^{(\dagger)}
\ldots U_{\bm{x}_n}^{(\dagger)}\ Z_\tau[U] \,.
\end{equation}
Here,
$U^{(\dagger)}$ is a generic notation for either $U$ or $U^\dagger$ 
(in any representation of
SU$(N)$), and
${\rm d}\mu(U)$ denotes the group invariant
measure  \cite{Miller,Zinn99}
(see also Appendix).
Eq.~(\ref{eq:expect-val}) is consistent with Balitsky's equations if
$Z_\tau[U]$ obeys the following equation:
\begin{equation}
\label{eq:RGdef}
  \partial_\tau Z_\tau[U]\,=\,\frac{1}{2}
\,\nabla_{\bm{x}}^a \, \chi^{a b}_{\bm{x} \bm{y}}[U] \,\nabla_{\bm{y}}^b\,
Z_\tau[U] ,
\end{equation}
where we employ a summation and integration convention for repeated
indices and transverse coordinates --- in the reminder of the text we
will only write integral signs where confusion might arise.
$\nabla_{\bm{x}}^a$ is a Lie derivative which generates translations on the
group manifold (see Appendix), 
\begin{equation}
   \label{eq:chidef}
\chi^{a b}_{\bm{x}\bm{y}}[U] \equiv
   \frac{1}{\pi} \int\!\! \frac{d^2 {\bm z}}{(2\pi)^2}\
  {\cal K}_{\bm{x} \bm{y}
     \bm{z}}
   \big[(1- U^\dagger_{\bm{x}}  U_{\bm{z}})(1 -  U^\dagger_{\bm{z}}
  U_{\bm{y}})\big]^{a b}
   \end{equation}
with $ U^{a b}$ in the {\it adjoint} representation, and
\begin{equation}
   \label{eq:Kdef}
   {\cal K}_{\bm{x} \bm{y}\bm{z}} \equiv
   \frac{(\bm{x}-\bm{z})\cdot(\bm{y}-\bm{z})}{%
     (\bm{x}-\bm{z})^2 (\bm{z}-\bm{y})^2}\,.
\end{equation}
$\chi$ is of the form ${\bf e}^\dagger {\bf e}$ and hence positive definite.
More precisely
\begin{equation}
\label{chidef}
\chi^{ab}_{\bm{x}\bm{y}}[U]\,=\,  \int d^2{\bm{z}}\, {\bf
e}^{ac,l}({\bm{x},\bm{z}})\, 
           {\bf e}^{bc,l}({\bm{y},\bm{z}}),
\end{equation}
which, according to our conventions,  we could have written shortly as
$\chi^{ab}_{\bm{x}\bm{y}}[U]={\bf e}^{ac,l}_{{\bm{x},\bm{z}}}
           {\bf e}^{bc,l}_{{\bm{y},\bm{z}}}$. The  
``square root'' factor in Eq.~(\ref{chidef}) is given by
\begin{equation}
\label{e}
e^{ab,l}({\bm{x},\bm{z}})
= \frac{1}{\sqrt{4 \pi^3}}\,\frac{(\bm{x}-\bm{z})^l}{(\bm{x}-\bm{z})^2}
\bigl(1-U^\dagger_{\bm{x}}U_{\bm{z}}\bigr)^{ab}.
\end{equation}
It is a matrix in  color indices
and a vector in transverse coordinates. With the help of
\begin{equation}
\label{eq:A-sigchi00}
  \sigma^a_{\bm{x}}[U]\,\equiv \,\frac{1}{2}\,
\nabla^b_{\bm{y}} \,\chi^{a b}_{\bm{x y}}[U],
\end{equation}
Eq.~(\ref{eq:RGdef}) takes the familiar  form of a Fokker-Planck
equation
\begin{equation}\label{eq:RGdef2} \partial_\tau  Z_\tau[U] \,=\, {1
  \over 2}\, \nabla^a_{\bm{x}}\nabla^b_{\bm{y}}\, \left( \chi^{ab}_{\bm{x} \bm{y}}[U]
\,Z_\tau[U]\right)\, - \, \nabla^a_{\bm{x}}\,\left(
\sigma^a_{\bm{x}}[U]\,Z_\tau[U]\right)\,, 
\end{equation}
albeit a functional one, involving fields with values on a curved
manifold.
As for all Fokker-Planck equations, there is an underlying random
process whose origin lies in the renormalisation group operation itself, and the
elimination of degrees of freedom which accompanies a change in $\tau$. In fact, this
aspect will be made the centerpiece of the formulation by recoding
Eq.~\eqref{eq:RGdef} as a random walk governed by  a Langevin equation.
However this short presentation does not make  the elements of the random
walks immediately visible; it does not either exhibit the important role
of the background field of the target in determining the properties of
the random walk. Those aspects are better seen with a picture of the
target wave function which utilizes the field $\alpha$ rather than $U$ as
the basic random variable, a description to which we will turn now.

\subsection{The $\alpha$-representation}
\label{sec:CGC}

The model of hadron wavefunction that we shall use rests on the
effective theory developed in Refs.  \cite{MV94, JKMW97, JKLW97, PI},
which we briefly summarize (see Ref. \cite{CARGESE} for details)

Because of the large rapidity gap between the dipole and the hadron,
the latter appears to the dipole as a Lorentz contracted colour source
for the small-x gluons to which it couples. We are then led to
separate the hadron constituents into {\em hard} ones with large
longitudinal momenta ($k^+\simge {\rm x}P^+$), and {\em soft} ones
with momenta $k^+\simle {\rm x}P^+$ (with $P^+=$
the total longitudinal momentum of the hadron).  The hard partons include the
valence quarks, as well as the partons created in the quantum
evolution down to longitudinal momenta ${\rm x}P^+$ and whose density
increases with $1/{\rm x}$. Together, these hard partons constitute a
high density system of colour charges which, during the interaction
with the dipole, can be regarded as frozen and constitue the source
${\rho}_a(x^-,{\bm{x}})$ of a classical field representing the soft
gluons.

This classical field is assumed to be the solution of the  Yang-Mills 
equations:
$
(D_{\nu} F^{\nu \mu})_a(x)\, =\, \delta^{\mu +} \rho_a(x^-,{\bm{x}})\,,
\label{cleq0}
$
where the source $\rho^a(x^-,{\bm{x}})$ is
time-independent, i.e.  independent of $x^+$. This assumption follows from the
observation that the internal dynamics of the hadron is frozen during 
the  relatively short
interaction time $\Delta x^+ \approx 2{\rm x}P^+/Q^2$  with the 
dipole. The classical field  has
just one independent component, which in the covariant  gauge 
$\partial_\mu {A}^\mu =0$ is the
component
$A^+_a(x^-,{\bm{x}})\equiv\alpha_a(x^-,{\bm{x}})$, with
$
- \partial^2_\perp \alpha_a(x^-,{\bm{x}})\,=\,{\rho}_a(x^-,{\bm{x}})\,.
$

The source $\rho_a$, or equivalently the classical field $\alpha_a$, is
treated as a random variable, with some gauge-invariant probability
density $W_\tau[\alpha_a]$, which depends upon $\tau\equiv\ln(1/{\rm x})$
\cite{JKMW97,JKLW97}.  In order to compute some observable ${\cal O}[\alpha]$
(to leading log accuracy), one first evaluates it with the classical
solution, and then average the result  with the weight function
$W_\tau[\alpha]$ :
\begin{equation}
\label{OBS}
\langle {\cal O}  \,\rangle_\tau = 
\int\,[{\rm d}\alpha]\,{\cal O}[\alpha] \,W_\tau[\alpha]\,.
\end{equation}
This peculiar averaging, somewhat reminiscent of the mathematical
description of amorphous materials like spin glasses, has been dubbed
{\em Colour Glass Condensate} \cite{PI}.  What is being done here is
analogous to the Born-Oppenheimer approximation: the original quantum
average is replaced by a classical average over the various possible
configurations of the ``frozen'' degrees of freedom.

The probability distribution $W_\tau[\alpha]$ is not known directly,
but its variation corresponding to freezing partons in the
rapidity window $(\tau,\tau+d\tau)$ can be computed. These partons
have longitudinal momenta $b\Lambda < k^+ <\Lambda$, where
$\Lambda$ is the momentum scale corresponding to $\tau$
(i.e., $\tau=\ln(P^+/\Lambda)$) and $d\tau=\ln(1/b)$. The elimination of these
degrees of freedom induces new correlations at the softer scale
$b\Lambda$, which, to the accuracy of interest, can be accounted for
by shifting the original field $\alpha^a(x^-,\bm{x})$ by a random quantity
$\alpha^a_\tau(\bm{x})\, d\tau$ with only 1-point and 2-point
correlation functions:
\begin{equation}
\label{a-fluct}
\langle
\alpha^a_\tau(\bm{x})\rangle_\Lambda\,,\qquad\qquad
\langle \alpha^a_\tau(\bm{x})\alpha^b_\tau(\bm{y})\rangle_\Lambda\,\,,
\end{equation}
where the subscript $\Lambda$ is here to remind us that the averages depend
on the background field $\alpha(x^-,{\bm{x}})$ at the scale $\Lambda$.

At this place we need to digress on a subtle aspect of the problem
which has to do with the longitudinal structure of the field. The
random field $\alpha^a_\tau(\bm{x})$ is a two dimensional field which is in
fact obtained by integrating over $x^-$ a fluctuation
$\delta\alpha^a_\tau(x^-,\bm{x})$:
\begin{equation}\label{integratedda}
\alpha^a_{\tau}({\bm{x}})\, d\tau\equiv \int dx^-
\delta\alpha^a_\tau(x^-,\bm{x}).
\end{equation}
It is shown in \cite{PI} that  the $x^-$ dependence is of the form:
\begin{equation}\label{FormF}
\delta\alpha^a_\tau
(x^-,\bm{x})\,\propto\,\theta(x^-)\,\frac{e^{-{\rm i}b\Lambda x^-}-
e^{-{\rm i}\Lambda x^-}}{x^-}\,,
\end{equation}
so that the integral over
$x^-$ in Eq.~(\ref{integratedda}) produces a factor $\ln(1/b)=d\tau$.
 Eq.~(\ref{FormF}) shows that
the shift in the classical field resulting from integrating out
gluons in the chosen layer of $k^+$ has support at $1/\Lambda \simle x^-
\simle 1/b\Lambda\/$, and therefore essentially no overlap with the original
field $\alpha^a(x^-,{\bm{x}})$ at the scale $\Lambda$
(whose support is $x^-\simle 1/\Lambda$).  This
is an important feature which allows us to reconstruct step by step
the Wilson lines involved in the calculation of high energy scattering
matrix elements. Thus, for instance:
\begin{equation}\label{Utau1}
U^\dagger_{\tau+d\tau}({\bm{x}})=\,{\rm e}^{{\rm i}g\int dx^-
\delta\alpha^a_\tau(x^-,\bm{x})}\,U^\dagger_{\tau}({\bm{x}})=
\,{\rm e}^{{\rm i}g d\tau
\alpha^a_\tau (\bm{x})}\,U^\dagger_{\tau}({\bm{x}})\,,
\end{equation}
since all the $x^-$ in $\delta\alpha^a_\tau(x^-,\bm{x})$ are larger 
than those contained in the
Wilson line $U^\dagger_{\tau}(\bm{x})$. This being recognized, it is 
not essential to keep trace
of the detailed $x^-$ dependance of the field exhibited in 
eq.~(\ref{FormF}), and we shall work
with integrated quantities such as $\alpha_\tau({\bm{x}})$: the 
ordering in $\tau$ takes
care of the initial ordering in
$x^{-}$. 
We are now in position to introduce the equation obeyed by the 
probability distribution
$W_\tau[\alpha]$\cite{PI} :
\begin{equation}\label{RGEA}
\partial_\tau  W_\tau[\alpha] \,=\,
  {1 \over 2} \frac{\delta^2}{\delta
\alpha^a_{\tau}({\bm{x}}) \delta \alpha^b_{\tau}({\bm{y}})}\,
\big(\chi^{ab}_{\bm{x y}} [\alpha] W_\tau[\alpha] \big) -
\frac{\delta}{\delta \alpha^a_{\tau}({\bm{x}})}\,
\big(\sigma^a_{\bm{x}}[\alpha] W_\tau[\alpha]\big) \,,
\end{equation}
where $\chi^{ab}_{\bm{x y}}[\alpha]$ and $\sigma^a_{\bm{x}}[\alpha]$ are given
by Eqs.~\eqref{eq:chidef} and~\eqref{eq:A-sigchi00} respectively, in which 
$ U$ and $ U^\dagger$ represent Wilson lines in the adjoint
representation, built from the classical field using Eq.~(\ref{Utau1}) iteratively
from $\tau_0$ to $\tau$.  
In Eq.~(\ref{RGEA}), the functional derivatives are taken
with respect to the random variable of the last rapidity bin, 
i.e. $\alpha_\tau$,
which, according to the previous discussion, 
is the only one involved in going from
$W_\tau[\alpha]$ to
$W_{\tau+d\tau}[\alpha]$.

\subsection{Path integral solutions and Langevin representations of the RG}
\label{sec:restults-sketch}

As emerges from the discussion in the previous subsection, the two 
evolution equations in the
$U$ and the $\alpha$ representations,  summarize a random process by 
which the Wilson lines are
built.  This process is more directly visible in the $\alpha$ 
representation: there we see
clearly that at each step in rapidity, the Wison line is multiplied 
by a random matrix close to
unity. The Wilson line at rapidity $\tau$  is given by
\begin{equation}
\label{U-path-alpha}
U^\dagger_{\tau}({\bm{x}})\,=\,{\rm T}\exp\bigg\{{\rm i}g\int_{0}^\tau  d
\eta\, \alpha_{\eta}({\bm{x}})\bigg\}U^\dagger_{0}({\bm{x}})\,,
\end{equation}
with, as already mentioned, the inital ordering in $x^-$ (cf. 
eq.~(\ref{Udagger})) replaced by an
ordering in $\tau$. With this explicit representation at hand, one 
has already indications
about the relation between the two approaches. Indeed,  the Lie 
derivative which enters the equation for
$Z[U]$, and which produces a ``translation'' of $U$, is equivalent 
when acting on $U$  such as
(\ref{U-path-alpha})
  to:
\begin{equation}
\nabla^a U^\dagger\,=\, 
\frac{\delta U^\dagger}{\delta \alpha_\tau^a},
\end{equation}
which is precisely the variation generated in the $\alpha$ 
representation. In fact one can write an
explicit formula relating the two probability distributions $ 
Z_\tau[U]$ and $W_\tau[\alpha]$:
\begin{equation}
\label{eq:U-alpha-0}
Z_\tau[U]\,=\,\int [{\rm d}\alpha]\,\,\delta_G\big(U^\dagger,\,
  {\rm T}\,{\rm e}^{{\rm i}g
\int_{0}^{\tau} {\rm d}\eta\,\alpha_{\eta}}
\big)\,W_\tau[\alpha]\,.
\end{equation}
This also immediately allows us to translate the equations for the two
into each other, giving complete equivalence of the approaches.

Thus the basic problem of the evolution at small x is that of a random
walk in a space of two dimensional fields with values in SU(N). The
group nature of the variable introduces several complication in the
random walk which will be dealt with in section 4. But the general
structure of the problem can be uncovered  without dealing with these
complications. This will be done in the next section.

\setcounter{equation}{0}
\section{Preliminaries: Brownian motion in flat space}
\label{sec:part-brownian}

As recalled in the previous sections, the evolution with rapidity of
scattering observables can be described by two distinct functional
equations. Both can be interpreted as diffusion equations, but they
have rather different structures and control the evolution of quite
different object: a probability distribution on group valued fields in
one case, a functional of gauge fields in the other. It is the purpose
of this section to clarify the deep relation between these two
formalisms by showing that they have counterparts in the description
of ordinary Brownian motion.

We shall consider the Brownian motion of a particle in a $D$-dimensional
Euclidean space.  As we shall see, the probability to find the particle
at point
${\bm{x}}$ at time $\tau$, which we call $P({\bm{x}},\tau)$, is related to
$Z_\tau[U]$, with the coordinate ${\bm{x}}$ corresponding to $U$.
Similarly one may introduce a probability functional $W_\tau[{\bm{v}}]$
over the random paths, where the paths are parametrized by the velocity
${\bm{v}}$ of the particle at each time step. This functional
$W_\tau[{\bm{v}}]$ is the analog of $W_\tau[\alpha]$. It is related to
$P({\bm{x}},\tau)$ by an equation analoguous to Eq.~(\ref{eq:U-alpha-0}):
\begin{equation}
\label{PW0}
P({\bm{x}},\tau)=\int [{\rm d}{\bm{v}}]\,
\delta^{(D)}({\bm{x}}-{\bm{x}}[{\bm{v}}])
W_\tau[{\bm{v}}]
\end{equation}
with
\begin{equation}\label{xv0}
{\bm{x}}[{\bm{v}}]={\bm{x}}_0+\int_0^T {\rm d}\tau' \,{\bm{v}}(\tau')
\end{equation}
with $T$ an arbitary time greater than $\tau$. Note that in the r.h.s. of
Eq.~(\ref{PW0}) all the time-dependence is carried by the weight function
$W_\tau[{\bm{v}}]$.
Of course, the paths which are physically
relevant for computing $P({\bm{x}},\tau)$
terminate at $\tau$: this information is  encoded in
the weight function $W_\tau[{\bm{v}}]$.

The probablility distribution $P({\bm{x}},\tau)$ obeys a Fokker-Planck
equation which, in analogy with Eq.~(\ref{eq:U-alpha-0})
satisfied by $Z_\tau[U]$, we choose of the form
\begin{equation}
   \label{eq:A-particle-FP}
   \Dot P({\bm{x}},\tau) = - H P({\bm{x}},\tau)
\end{equation}
where the dot denotes the time derivative, $H$ is the
following``hamiltonian'':
\be\label{eq:Ham-varia}
  H\,\equiv \, -\partial^\alpha\Big[\partial^\beta
\frac{1}{2}\chi^{\alpha\beta}({\bm{x}})-\sigma^\alpha({\bm{x}})\Big],\ee
(with $\partial^\alpha=\del/\del x^\alpha$)  and
$\chi^{\alpha\beta}({\bm{x}})$ is of the form:
\begin{eqnarray}
\label{eq:Brownian-chi}
\chi^{\alpha\beta}({\bm{x}})\,\equiv\,
e^{\alpha\gamma}({\bm{x}})e^{\beta\gamma}({\bm{x}}).
\end{eqnarray}
Thus $\chi^{\alpha\beta}({\bm{x}})$ is symmetric
and positive semi-definite, and therefore invertible.
We shall also assume the following relation, analogous to
Eq.~(\ref{eq:A-sigchi00}):
\begin{equation}
\label{eq:A-sigchi}
  \sigma^\alpha({\bm{x}})\,=\,\frac{1}{2}\,
\partial^\beta \,\chi^{\alpha\beta}({\bm{x}})
\end{equation}
  which, in particular, guarantees that $H$ in Eq.~(\ref{eq:Ham-varia})
is hermitian. The equation satisfied by  $W_\tau$ will be found to be:
\begin{equation}
\label{eq:W-velocity-equation-0}
\partial_\tau  W_\tau[{\bm{v}}]\,=\,
\frac{\delta}{\delta v^\alpha_\tau}\left[\frac{1}{2}
\frac{\delta}{\delta v^\beta_\tau}\Big(\chi^{\alpha\beta}({\bm{x}})
W_\tau[{\bm{v}}]\Big) - \sigma^\alpha({\bm{x}})W_\tau[{\bm{v}}]\right].
\end{equation}

Note that the  Fokker-Planck equation
(\ref{eq:A-particle-FP})--(\ref{eq:Ham-varia}) for
$P({\bm{x}},\tau)$ follows indeed
from Eq.~(\ref{eq:W-velocity-equation-0}) for $W_\tau[{\bm{v}}]$
together with the relation (\ref{PW0}), as
can be checked by  performing some integrations by parts with respect
to ${\bm{v}}$ in Eq.~(\ref{PW0}) and then using the identity:
\begin{equation}
\frac{\delta}{\delta v^\alpha_\tau}
\,\delta ({\bm{x}}-{\bm{x}}[{\bm{v}}])\,=\,-\partial^\alpha
\delta ({\bm{x}}-{\bm{x}}[{\bm{v}}]).
\end{equation}
Underlying all these equations is an elementary random process which
controls the random walk. This elementary process is best seen at the
level of a Langevin equation, from which in fact all equations can be
derived. We shall make this Langevin equation the starting point of our
discussion.

\subsection{Langevin equation and path integral}
\label{sec:Langevin}

Consider then a particle undergoing a random walk in a $D$-dimensional
Euclidean space. Its coordinates  $x^\alpha$ ($\alpha, \beta
=1,\cdots,D$) obey the following Langevin equation:
\begin{equation}
\label{eq:Langevin-formal}
\dot x^\alpha(\tau)\,=\,\sigma^\alpha({\bm{x}}) + e^{\alpha\beta}({\bm{x}})
\,\nu^\beta(\tau),
\end{equation}
where the dot denotes time derivative, and $\sigma^\alpha({\bm{x}})$ and
$e^{\alpha\beta}({\bm{x}})$ are given regular functions of the coordinates
characterizing the medium in which the particle propagates; finally,
$\nu^\alpha(\tau)$ is a Gaussian white noise (see below). We shall also use the
more compact notation:
\begin{eqnarray}
\label{eq:Langevin-formal2}
\dot {\bm{x}}(\tau)\,=\,\bm{\sigma}({\bm{x}}) + e({\bm{x}})
\,\bm{\nu}(\tau),
\end{eqnarray}
where ${\bm{x}}, {\bm{\sigma}}$ and $\bm{\nu}$ are $D$- dimensional
vectors and $e$ is a $D\times D$ matrix.

Strictly speaking, Eq.~(\ref{eq:Langevin-formal}) is formal (the
trajectory ${\bm{x}}(\tau)$ of the random walk is not differentiable) and
gets meaningful only with a discretization prescription. Here, we
shall focus on the following discretization:
\begin{eqnarray}
\label{eq:Langevin-discrete}
   \frac{{\bm{x}}_i- {\bm{x}}_{i-1}}{\epsilon}\,=\,
\bm{\sigma}_{i-1} + e_{i-1}\,\bm{\nu}_i,
\end{eqnarray}
and only briefly mention later other possible choices
(see also Refs. \cite{ZJ,Arnold}).
In Eq.~(\ref{eq:Langevin-discrete}), $\epsilon$ is the length of the
time step, ${\bm{x}}_i\equiv{\bm{x}}(\tau_i)$ with $\tau_i=i\epsilon$,
$i=0\,,1\,,\cdots n$, and ${\bm{x}}_0\equiv {\bm{x}}(\tau_0)$ is a given
initial condition. We have also introduced the simplified notation
$\bm{\sigma}_{i}\equiv\bm{\sigma}({{\bm{x}}_i})$, $ e_i\equiv e({{\bm{x}}_i})$,
to be used throughout.

The $\bm{\nu}_i\,'$s are Gaussian random variables with zero expectation
value and local 2-point correlations:
\begin{equation}
\label{eq:noise-corr}
\langle \nu^\alpha_i \nu^\beta_j \rangle\,=\,\frac{1}{\epsilon}
\,\delta^{\alpha\beta}
\delta_{ij}.
\end{equation}
In other words, the probability law for the noise variables $\bm{\nu}_i$ is
a normalized Gaussian distribution:
\begin{equation}
\label{eq:prob-noise}
  d{\cal P}(\bm{\nu}_i)\equiv\,
\left(\frac{\epsilon}{2\pi}\right)^{D/2}
\,{\rm e}^{-\frac{\epsilon}{2}\,
\bm{\nu}_i\cdot\bm{\nu}_i}\,{d^D\bm{\nu}_i}\,,
\end{equation}
the same for each time step, and for $n$ time steps it is the product
of $n$ such distributions.
  
The particular discretization (\ref{eq:Langevin-discrete}) of the
Langevin equation is such that, for fixed ${\bm{x}}_{i-1}$,
${\bm{x}}_i$ is a random variable linearly related to $\bm{\nu}_i$.
This makes it easy to calculate the probability
$P_\epsilon({\bm{x}}_i|{\bm{x}}_{i-1})$ that the random walk brings the
particle from position ${\bm{x}}_{i-1}$ to ${\bm{x}}_i$ in one time
step, and this for any value of $\epsilon$. (We shall often refer to
quantities such as $P_\epsilon({\bm{x}}_i|{\bm{x}}_{i-1})$ as ``elementary
propagators''.) We have:
\begin{equation}
\label{eq:P-prop-step1}
P_\epsilon({\bm{x}}_i|{\bm{x}}_{i-1}) = \int d{\cal P}(\bm{\nu}_i)\,
\delta^D\big({\bm{x}}_i-{\bm{x}}_{i-1}-\epsilon(\bm{\sigma}_{i-1}+e_{i-1}\bm{\nu}_i)\big).
\end{equation}
The integral over $\bm{\nu}_i$ is easily performed, and yields:
\begin{equation}
\label{eq:P-prop-epsilon}
P_\epsilon({\bm{x}}_i|{\bm{x}}_{i-1})=
\frac{(\det\chi_{i-1})^{-1/2}}{(2\pi\epsilon)^{D/2}}\,
  \,{\rm e}^{-\frac{1}{2\epsilon}\big({\bm{x}}_i-{\bm{x}}_{i-1}-\epsilon
\bm{\sigma}_{i-1}\big) \,\chi^{-1}_{i-1}\,({\bm{x}}_i
-{\bm{x}}_{i-1}-\epsilon \bm{\sigma}_{i-1})}
\, ,
\end{equation}
where $\chi$ is obtained from $e$ via Eq.~(\ref{eq:Brownian-chi}). 
Thus, for fixed ${\bm{x}}_{i-1}$, the probability to find the particle
at position ${\bm{x}}_i$ is a gaussian centered around
${\bm{x}}_i={\bm{x}}_{i-1}-\epsilon\bm{\sigma}_{i-1}$, with a width
$\epsilon\chi_{i-1}$.

The coordinate ${\bm{x}}_n$ of the particle after $n$ time step is 
also a random variable. The
average of any function of ${\bm{x}}_n$, say $f({\bm{x}}_n)$, can be  obtained by
first solving Eq.~(\ref{eq:Langevin-discrete}) for a given realization
$\{\bm{\nu}_1,\bm{\nu}_2,\cdots,\bm{\nu}_n\}$ of the random variables, and then
averaging  over all such realizations. Equivalently, one may first
determine the
probability  density
$P({\bm{x}},\tau)$ to find the particle at point ${\bm{x}}$ at time
$\tau=n\epsilon$ knowing that it was at ${\bm{x}}_0$ at time $\tau_0=0$.
The average of $f({\bm{x}})$ may then be obtained as
\be
\langle f({\bm{x}}_n)\rangle=\langle f({\bm{x}})\rangle_\tau= \int
{\rm d}^D{\bm{x}}\, f({\bm{x}})\, P({\bm{x}},\tau)
\ee
The probability $P({\bm{x}},\tau)$ is given by:
\begin{eqnarray}
\label{eq:P-prop}
P({\bm{x}},\tau)\,\equiv\,\int \prod_{i=1}^n d
{\cal P}(\bm{\nu}_i)\,\,\delta^D\big(
{\bm{x}}-{\bm{x}}_n[\bm{\nu}]\big),
\end{eqnarray}
where ${\bm{x}}_n[\bm{\nu}]$ is the solution of
Eq.~(\ref{eq:Langevin-discrete}) for $n$ time steps, and depends
therefore on all the $\bm{\nu}_i$'s with $i\leq n$.  Using
repeatedly Eq.~(\ref{eq:P-prop-step1}) one can easily show that  the
integral above can be rewritten as:
\be\label{eq:recurr}
P({\bm{x}},\tau)=\int\prod\limits_{i=1}^{n-1}d^D{\bm{x}}_{i}\,\,
P_\epsilon({\bm{x}}|{\bm{x}}_{n-1})\,
\prod\limits_{i=1}^{n-1}P_\epsilon({\bm{x}}_i|{\bm{x}}_{i-1}).
\ee
By inserting in this expression the formula (\ref{eq:P-prop-epsilon})
for the elementary propagator, we finally get:
\be\label{eq:part-time}
P({\bm{x}},\tau)=
\int [d{\bm{x}}] \,  \delta({\bm{x}}-{\bm{x}_n})\,{\rm e}^{-{\cal A}[{\bm{x}}]}\,,
\ee
with
\begin{equation}
[d{\bm{x}}]\equiv \prod\limits_{i=1}^{n}\frac{d^D{\bm{x}}_{i}}{(2\pi\epsilon)^{D/2}}\,
\,\prod\limits_{i=1}^{n} (\det\chi_{i-1})^{-1/2} \,,
\end{equation}
and
\begin{equation}
{\cal A}[{\bm{x}}]\,\equiv\,\frac{\epsilon}{2}\sum_{i=1}^n\Big(
\frac{{\bm{x}}_i-{\bm{x}}_{i-1}}{\epsilon}-\bm{\sigma}_{i-1}\Big)
\chi^{-1}_{i-1}
\Big(
\frac{{\bm{x}}_i-{\bm{x}}_{i-1}}{\epsilon}-\bm{\sigma}_{i-1}\Big)
\end{equation}
The average $\langle f({\bm{x}})\rangle_\tau$ is then easily obtained:
\begin{equation}
\langle f({\bm{x}})\rangle_\tau=\int {\rm d}^D{\bm{x}}\, f({\bm{x}})\, P({\bm{x}},\tau)=
\int [d{\bm{x}}] \,\, f({\bm{x}}_n)\, {\rm e}^{-{\cal A}[{\bm{x}}]}
\end{equation}
An alternative expression for $P({\bm{x}},\tau)$ is obtained by
integrating over auxiliary momentum variables which allows one to get rid
of the determinant
$\det\chi_{i-1}$ in the measure and to reexpress ${\cal A}$ in terms of
the matrix $\chi_{i-1}$ rather than its inverse:
\begin{equation}\label{PL}
P({\bm{x}},\tau)=\int\prod\limits_{j=1}^{n}\bigg[d^D{\bm{x}}_{j}\frac
{d^D{\bm{p}}_{j}}{(2\pi)^D}\bigg] \delta({\bm{x}}-{\bm{x}_n})
\exp\left\{-\sum_{j=1}^n[ {\rm i}{\bm{p}}_{j}\cdot({\bm{x}}_j-{\bm{x}}_{j-1}) 
  +\epsilon{\cal L}_j]\right\}
\end{equation}
with
\begin{equation}
{\cal L}_j[{\bm{x}},{\bm{p}}]\,\equiv\,
\frac{1}{2}\,{\bm{p}}_j\cdot \chi_{j-1}\cdot {\bm{p}}_j
- {\rm i}{\bm{p}}_j\cdot \bm{\sigma}_{j-1}.
\end{equation}

The formulae (\ref{eq:part-time}) and (\ref{PL}) express the
probability $P({\bm{x}},\tau)$ as a sum over paths specified by the set
of coordinates $\{{\bm{x}}_0,\,{\bm{x}}_1,\,\cdots {\bm{x}}_{n-1},
{\bm{x}}_{n}\}$, where ${\bm{x}}_0$ is a given initial condition while
all the other coordinates are random variables. A noteworthy feature
of this expression is that, at step $i$, the functions $\chi({\bm{x}})$
and $\sigma({\bm{x}})$ have to be evaluated at the point ${\bm{x}}_{i-1}$
attained in the preceding step. This property is of course intimately
related to the special discretization of the Langevin equation,
Eq.~(\ref{eq:Langevin-discrete}). One could naively expect that, in
the limit $\epsilon \to 0$, ${\bm{x}}_{i-1}$ can be replaced by ${\bm{x}}_i$ in
$\chi$ or $\sigma$. This is not so however, and it is important to keep the
arguments of $\chi$ and $\sigma$ as written in Eqs.~(\ref{eq:part-time}) or
(\ref{PL}). This is because the typical paths contributing to the
integral are such that $|{\bm{x}}_i-{\bm{x}}_{i-1}|\sim \sqrt{\epsilon}$.  Thus,
the replacement of ${\bm{x}}_{i-1}$ by ${\bm{x}}_i$ within $\chi$ or $\sigma$
modifies the exponent in the integrand of these formulae by terms of
the same order of magnitude as the terms that have been kept.
Therefore, writing Eq.~(\ref{PL}) in continuum notations, i.e., as 
\begin{equation}\label{eq:cont-A}
  P({\bm{x}},\tau) =
\int_{{\bm{x}}_{0}}^{{\bm{x}}}
\![{\rm d}{\bm{x}}] \int\![{\rm d}{\bm{p}}]  \exp\left\{
\int_{0}^\tau\!{\rm d}\eta\big[-{\rm i}{\bm{p}}\cdot \big(\Dot {\bm{x}}
-\bm{\sigma} ({\bm{x}} )\big) -
    \frac{1}{2}\, {\bm{p}}\cdot\chi({\bm{x}})\cdot{\bm{p}}
     \big]\right\},
\end{equation}
is ambiguous.

\subsection{The Fokker-Planck equation}
   \label{sec:FP-eq}

The path integral
in Eq.~(\ref{eq:part-time}) summarizes in a compact formula the 
averages performed at each step
of the Langevin process. In the limit
$\epsilon\to 0$, it is possible to obtain  for  $P({\bm{x}},\tau)$ a 
differential equation which
involves only the local properties of the medium in which the 
particle propagate, i.e. the
functions $\chi$ and $\sigma$, and with no explicit reference to the specific
elementary random process. This equation is a Fokker-Planck equation.

To obtain this equation we note first that Eq.~(\ref{eq:recurr}) allow
us to write: \be\label{P-exp} P({\bm{x},\tau})= \int d^D{\bm{x}}_{n-1}
\,P_\epsilon({\bm{x}}|{\bm{x}}_{n-1})\,P({\bm{x}}_{n-1},\tau_{n-1}).  \ee In
order to calculate the integral over ${\bm{x}}_{n-1}$, we use the fact
that the propagator $P_\epsilon({\bm{x}}|{\bm{x}}_{n-1})$, considered as a
function of ${\bm{x}}_{n-1}$, is a gaussian which, when $\epsilon\to 0$, is
strongly peaked near ${\bm{x}}$ (see Eq.~(\ref{eq:P-prop-epsilon}));
it follows that, when $\epsilon$ is small, one can get a good approximation
of the integrand of Eq.~(\ref{P-exp}) by expanding
$P({\bm{x}}_{n-1},\tau_{n-1})$ in the vicinity of ${\bm{x}}$.  To do that
systematically, it is in fact easier to go back to
Eq.~(\ref{eq:P-prop-step1}), and expand the $\delta$-constraint, keeping  all terms that will
contribute to ${\cal O}(\epsilon)$ after averaging over the noise:
\begin{align}
  \label{eq:delta_expansion}
\delta^D\big( & {\bm{x}}-{\bm{x}}_{n-1}
-\epsilon(\bm{\sigma}_{n-1}+e_{n-1}\bm{\nu}_n)\big) =
 \,{\rm
e}^{-(\bm{\sigma}_{n-1}+e_{n-1}\bm{\nu}_n){\cdot}\partial}\delta^D\big({\bm{x}}-{\bm{x}}_{n-1}\big)
\nonumber \\
 \approx &\Bigl(1-(\bm{\sigma}_{n-1}+e_{n-1}\bm{\nu}){\cdot}\partial+\tfrac{1}{2} ((\bm{\sigma}_{n-1}+e_{n-1}\bm{\nu}){\cdot}\partial)^2
\Bigr)\delta^D\big({\bm{x}}-{\bm{x}}_{n-1}\big)\ ,
\end{align}
where $\partial_\alpha=\del/\del x^\alpha$.
Averaging over $\nu_n$ leaves us with the expression
\begin{equation}\label{expansion}
P_\epsilon({\bm{x}}|{\bm{x}}_{n-1})=\left(
1 - \epsilon\sigma_{n-1}^\alpha \partial_\alpha
+\frac{\epsilon}{2}\,\chi_{n-1}^{\alpha\beta}
\partial_\alpha \partial_\beta  \right)
\delta^D({\bm{x}}-{\bm{x}}_{n-1})+
{\cal O}
(\epsilon^{3/2}) \ .
\end{equation}
  The first term obtained after plugging this formula into
Eq.~(\ref{P-exp}) is simply:
\be
P({\bm{x}},\tau_{n-1})
\,=\,P({\bm{x}},\tau-\epsilon)\,\approx\,P({\bm{x}},\tau) - \epsilon\dot
P({\bm{x}},\tau),\ee
where $\dot P({\bm{x}},\tau)\equiv \partial_\tau P({\bm{x}},\tau)$. In the
remaining terms, we use identities like
\begin{equation}
\sigma^\alpha({\bm{x}}_{n-1})\partial_\alpha\,
\delta^D({\bm{x}}-{\bm{x}}_{n-1})\,P({\bm{x}}_{n-1},\tau_{n-1}) = 
\partial_\alpha \big(
\delta^D({\bm{x}}-{\bm{x}}_{n-1})\sigma^\alpha({\bm{x}})P({\bm{x}},\tau_{n-1})\big), 
\end{equation}
to deduce that, in the limit $\epsilon\to
0$,
$P({\bm{x}},\tau)$ obeys
  the following equation:
\begin{eqnarray}
\label{eq:particle-FP-0}
    \Dot P({\bm{x}},\tau) = \partial^\alpha\Big[\partial^\beta
\Big(\frac{1}{2}\chi^{\alpha\beta}({\bm{x}})P({\bm{x}},\tau)\Big)-
\sigma^\alpha(x)P({\bm{x}},\tau)\Big].
\end{eqnarray}

\subsection{Alternative discretizations}
\label{sec:same-phys-diff}

At this point, one could reverse the logics and consider the
Fokker-Planck equation as our starting point. Then, path integral
representations for the solution $P({\bm{x}},\tau)$ could be
constructed in a standard way, for instance by exploiting the formal
analogy between Eq.~(\ref{eq:particle-FP-0}) and the Schr{\"o}dinger
equation in imaginary time. Because the coefficients $\chi({\bm{x}})$
and $\sigma({\bm{x}})$ in this equation are coordinate dependent,
different orderings of the derivatives within the Hamiltonian
(\ref{eq:Ham-varia}) will lead to different path-integral
representations (possibly involving derivatives of $\chi$ and
$\sigma$), which become however equivalent in the limit $\epsilon\to 0$.

To illustrate this, let us consider three alternative expressions of
the Hamiltonian (\ref{eq:Ham-varia}) which are obtained by
successively commuting the derivatives with $\chi$, and using
Eq.~(\ref{eq:A-sigchi}) to simplify the results:
\begin{subequations}
   \label{eq:Ham2}
\begin{align}
\label{eq:Ham2a}
  H\,= & -\left(\tfrac{1}{2}\,\partial^\alpha\partial^\beta
\chi^{\alpha\beta}({\bm{x}})
-\partial^\alpha\sigma^\alpha({\bm{x}})\right)
    \\
\label{eq:Ham2b}
\,= &-\left(\tfrac{1}{2}\,\chi^{\alpha\beta}({\bm{x}})\partial^\alpha\partial^\beta
+\sigma^\alpha({\bm{x}})\partial^\alpha\right)
\\
\label{eq:Ham2c}
\,= &-\frac{1}{8}\,\{\partial^\alpha,\{\partial^\beta,
\chi^{\alpha\beta}({\bm{x}})\}\} + \frac{1}{8}\,(\partial^\alpha
\partial^\beta \chi^{\alpha\beta}({\bm{x}})).
  \end{align}  \end{subequations}
The form in Eq.~(\ref{eq:Ham2a}) is what naturally came out of our original
Langevin discretization, the other two represent different but clearly
equivalent orderings of momentum and coordinate operators; the term
involving a double anti-commutator in Eq.~(\ref{eq:Ham2c}) is in the
so-called  Weyl-ordered form.  By using the last two expressions,
Eqs.~(\ref{eq:Ham2b}) and (\ref{eq:Ham2c}), within a canonical
construction of the path-integral, one obtains expressions similar to
Eq.~(\ref{PL}), but with $\sigma$ and $\chi$ evaluated at
${\bm{x}}_{i}$ (for Eq.~(\ref{eq:Ham2b})) and, respectively, at the
midpoint $\bar {\bm{x}}_i\equiv ({\bm{x}}_i+{\bm{x}}_{i-1})/2$ (for
Eq.~(\ref{eq:Ham2c})). The only difference with respect to
Eq.~(\ref{PL}) consists in the replacement of ${\cal L}_i$ in that
equation by the alternative expressions
\begin{subequations}
   \label{eq:part-advanced-Weyl}
\begin{align}
\label{eq:part-advanced}
{\cal L}_i = \,&\frac{1}{2}\,
\chi^{\alpha \beta}({\bm{x}}_i)     p_i^\alpha p_i^\beta
    + i  p_i^\alpha \sigma^\alpha({\bm{x}}_i)
      \\
\label{eq:part-Weyl}
      = \,&\chi^{\alpha \beta}(\Bar {\bm{x}}_i) p_{i}^{\alpha} p_{i}^{\beta}
     +\frac{1}{8}\partial^\alpha \partial^\beta \chi^{\alpha \beta}(\Bar
{\bm{x}}_i).
  \end{align}  \end{subequations}
It can be verified that the expressions of $P({\bm{x}},\tau)$ obtained from
Eqs.~(\ref{eq:part-advanced-Weyl}) satisfy indeed the  Fokker-Planck equation
(\ref{eq:particle-FP-0}) in the continuum limit.

Taking this one step further one can translate these differently
discretized path integrals into differently discretized Langevin
systems. Let us illustrate this for the case of Eq.~\eqref{eq:Ham2b}
with \eqref{eq:part-advanced} as its discretization prescription for
the path integral. One can verify that its 
elementary propagator can be written as the following integral over
 an auxiliary random noise
\begin{equation}
\label{eq:P-prop-step-advanced}
P_\epsilon({\bm{x}}_i|{\bm{x}}_{i-1}) = \int d{\cal P}(\bm{\nu}_i)\,
\delta^D\big(
{\bm{x}}_i-{\bm{x}}_{i-1}-\epsilon(-\bm{\sigma}_{i}+e_{i}\bm{\nu}_i)
\big).
\end{equation}
The $\delta$ function in this equation imposes the proper discretization of
the corresponding Langevin equation. Note that this has
$\bm{\sigma}$ and
$e$ evaluated at
${\bm x}_i$ instead of ${\bm x}_{i-1}$ and the {\em sign} of the
$\bm{\sigma}$ term is changed compared to
Eq.~\eqref{eq:Langevin-discrete}. For the Weyl ordered form, the
$\bm{\sigma}$ term would be completely absent but additional terms would
arise from eliminating the contribution
$e^{-\epsilon\frac{1}{8}\partial^\alpha \partial^\beta \chi^{\alpha \beta}}$
 from the measure in order to recover the structure of Eq.~(\ref{eq:P-prop-step-advanced}) necessary to read off the Langevin equation.

From here it is evident that continuous notations, which are not
able to keep track of details of coordinate dependence, are ambiguous,
both for path-integrals and Langevin descriptions.  The naive
continuum notation for the Langevin equation corresponding to
Eq.~\eqref{eq:P-prop-step-advanced} would leave us with the sign of
the $\bm{\sigma}$ term reversed compared to that in 
Eq.~\eqref{eq:P-prop-step1},
while ``forgetting'' the mid point prescription of the Weyl ordered
version would lead to yet another form. The same holds for
path-integral expressions involving the differently ordered ${\cal L}$
given above.

The one common denominator of all these equivalent, but differently
discretized, path-integrals and Langevin systems is the fact that they
correspond to one common Fokker-Planck equation which is perfectly
meaningful in continuum notation. 

\subsection{The velocity representation}
\label{sec:FP-velocity}

Until now we have focused on the ${\bm{x}}_n$ as the important random
variable. There are situations however where one needs to calculate
observables which depend on the  entire path, and not on the end point alone.
  Thus, we are led to consider
distribution functions over the paths.

To proceed with the construction of such a
  distribution, it is useful
to perform a change of variables, and to better specify the paths.

Let us then consider as new random variables the
  velocities:
\begin{eqnarray}
\label{eq:Langevin-velocity}
{\bm{v}}_i\,\equiv\,\frac{{\bm{x}}_i- {\bm{x}}_{i-1}}{\epsilon}
\,=\,\bm{\sigma}_{i-1} +\,
   e_{i-1}\bm{\nu}_i\,,
\end{eqnarray}
for $i=1,2, \cdots, N$,
where $T=N\epsilon$ is a fixed time much larger than all times of
interest. The coordinates of a path are easily expressed in terms of these:
\begin{eqnarray}
\label{eq:x-velocity}
{\bm{x}}_{i-1}\,=\,{\bm{x}}_0 + {\epsilon}\sum_{j=1}^{i-1}
{\bm{v}}_j\,
\equiv\,{\bm{x}}_{i-1}[{\bm{v}}]\,.
\end{eqnarray}
Note that this change of variables is not a linear
transformation  since, as is evident from Eq.~(\ref{eq:x-velocity}),  the
coefficients ${\bm \sigma}_{i-1}={\bm \sigma}({\bm{x}}_{i-1})$ and
$e_{i-1}=e({\bm{x}}_{i-1})$ in Eq.~(\ref{eq:Langevin-velocity}) for
${\bm{v}}_i$ involve the velocities
${\bm{v}}_j$ at all the steps $j=1,2,\,,\cdots i-1$. However,
$\sigma_{i-1}$ and $e_{i-1}$ do not depend upon ${\bm{v}}_i$ itself, so
that the  corresponding Jacobien is rather simple, being the
determinant of a triangular matrix.
Accordingly, the probability
distribution for the ${\bm{v}}_i$'s  is immediately obtained from that of
the ${\bm{x}}_i$'s in  Eq.~(\ref{eq:part-time}):
\be\label{eq:prob-velocity}
d{\cal P}[{\bm{v}}]\,=\, [d{\bm{v}}] \,  {\rm e}^{-{\cal A}[{\bm{v}}]}\,,
\ee
with
\be
[d{\bm{v}}]\equiv \prod_{i=1}^N\left(\frac{\epsilon}{2\pi}\right)^{D/2}
d^D{\bm{v}}_i\,\big({\rm det} \,\chi_{i-1}\big)^{-1/2} \,,
\ee
and
\be
{\cal A}[{\bm{v}}]\equiv\,\frac{\epsilon}{2}\,\sum_{i=1}^N
\big({\bm{v}}_i-\bm{\sigma}_{i-1}\big) \,\chi_{i-1}^{-1}\,
\big({\bm{v}}_i-\bm{\sigma}_{i-1}\big)\,.
\ee

We can write $d{\cal P}[{\bm{v}}]$ in the convenient factorized
form:
\begin{equation}\label{Pvfact}
d{\cal P}[{\bm{v}}] \equiv  \prod_{i=1}^N d^D{\bm{v}}_i\,{\cal P}_i[{\bm{v}}].
\end{equation}
Although this factorization property is a major advantage of the
velocity representation, it should be kept in mind that the notation
is deceiving since the different factors do not correspond to
independent probabilities: the probability density ${\cal
  P}_i[{\bm{v}}]$ for the velocity at step $i$ depends also upon the
velocities ${\bm{v}}_1$, ${\bm{v}}_2,\,\cdots {\bm{v}}_{i-1}$. Giving this,
a convenient way to check the normalization condition $\int d{\cal
  P}[{\bm{v}}]=1$ (or to do any other similar calculation) is to
perform the integrations over the ${\bm{v}}_i$'s in decreasing order
of $i$, from $i=N$ down to $i=1$ (notice that
Eq.~(\ref{eq:prob-velocity}) is truly a Gaussian in ${\bm{v}}_N$).

Let us turn now to the specification of the paths. The paths which
contribute to $P({\bm{x}},\tau)$ have a number $n=\tau/\epsilon$ of time steps, and
the probability $w_n({\bm{v}}_1,\cdots,{\bm{v}}_n)$ that a given such path
is realized in the random walk is given by Eq.~(\ref{Pvfact}) in which
$N$ is replaced by $n$, that is
\begin{equation}
w_n({\bm{v}}_1,\cdots,{\bm{v}}_n)= \prod_{i=1}^n d^D{\bm{v}}_i\,{\cal
P}_i[{\bm{v}}].
\end{equation}
Now we want to be able to compare such distributions for different
numbers of time steps, say $n$ and $n+1$, and eventually obtain a
differential equation in the limit of infinitesimal time steps.
Clearly one cannot directly compare $w_n$ and $w_{n+1}$ since these
have different numbers of variables. To allow such a comparison, one
possibility is to consider $w_n$ as a function of $n+1$ variables in
which the last one, ${\bm{v}}_{n+1}$, is set equal to zero. A
normalized distribution which achieves this goal is
$w_n({\bm{v}}_1,\cdots,{\bm{v}}_n)\delta({\bm{v}}_{n+1})$. More generally, we
shall consider paths with a fixed number $N$ of time steps such that
the velocities associated with the last $N-n$ steps vanish, while the
velocities of the first $n$ steps are random variables chosen with the
distribution given above.  That is, we define the following
probability distribution:
\begin{equation}
\label{eq:W-velocity-discrete0}
W_n[{\bm{v}}]\,\equiv\,
\prod_{i=1}^n\,{\cal P}_i[{\bm{v}}]\prod_{j=n+1}^N \delta^D({\bm{v}}_j)
\,.
\end{equation}
This distribution depends on $n$, that is on time, and satisfies an
evolution equation that we now determine. To this aim, we use
Eq.~(\ref{eq:W-velocity-discrete0}) to write:
\begin{equation}\label{Wnn-1}
  W_n[{\bm{v}}]-W_{n-1}[{\bm{v}}]\,=\,\left(
\prod_{i=1}^{n-1}\,{\cal P}_i[{\bm{v}}]\right)\left(\prod_{j=n+1}^N
\delta^D({\bm{v}}_j)\right)\,
\Big({\cal P}_n[{\bm{v}}]- \delta^D({\bm{v}}_n)\Big)\,.
\end{equation}
Then we note that ${\cal P}_n[{\bm{v}}]$, considered as a function of
$\epsilon{\bm{v}}_n$, is a gaussian peaked at ${\bm{v}}_n=\sigma_{n-1}$, and
proceed as for Eq.~(\ref{expansion}) to write:
\begin{eqnarray}
\label{expansion2}
  {\cal P}_n[{\bm{v}}]\approx\left(1-\sigma_{n-1}^\alpha\frac{\partial}{\partial
v_n^\alpha}\,+\,
\frac{1}{2\epsilon}\,\chi_{n-1}^{\alpha\beta}\frac{\partial^2}{\partial
v^\alpha_n\partial v^\beta_n}
\right)\delta^D({\bm{v}}_n)\,.
\end{eqnarray}
In writing such an expansion, we anticipated on the fact that the
relevant variable is actually $\epsilon {\bm{v}}$ rather than ${\bm{v}}$: it
is in the variable $\epsilon {\bm{v}}_n$ that the width of the gaussian is of
order $\epsilon$; furthermore, as we shall see shortly, $\epsilon {\bm{v}}_n$ is the
variable which is needed to turn $W_n[{\bm{v}}]$ into a functional of
the continuous function ${\bm{v}}(t)$.

Using  the expansion (\ref{expansion2}) in Eq.~(\ref{Wnn-1}) one
easily obtains:
\begin{eqnarray}
\label{eq:Wn-n-1}
   W_n[{\bm{v}}]-W_{n-1}[{\bm{v}}]\,=\,
\left(
\frac{1}{2\epsilon}\,\frac{\partial^2}{\partial v_n^\alpha\partial
v_n^\beta}\,\chi_{n-1}^{\alpha\beta}
\,-\,\frac{\partial}{\partial v_n^\alpha}\,\sigma_{n-1}^\alpha
\right)W_{n-1}[{\bm{v}}]\,.
\end{eqnarray}
Recall that $\chi_{n-1}$ and ${\bm \sigma}_{n-1}$ involve only the
velocities
${\bm{v}}_i$ with
$i\leq n-1$, cf. Eq.~(\ref{eq:x-velocity}).
However, since the function $W_{n-1}[{\bm{v}}]$  on the r.h.s.
is proportional to $\prod_{i\geq n}\delta^D({\bm{v}}_i)$ one may
replace ${\bm{x}}_{n-1}$ by $ {\bm{x}}[{\bm{v}}]$ in $\chi_{n-1}$ and
${\bm \sigma}_{n-1}$, with
\begin{eqnarray}
\label{eq:x-velocity-int-0}
{\bm{x}}[{\bm{v}}]\,\equiv\,
{\bm{x}}_0 +\epsilon\sum_{i=1}^N {\bm{v}}_i\,,
\end{eqnarray}
and consequently $\chi_{n-1}$ by $ \chi({\bm{x}})$ and ${\bm
\sigma}_{n-1}$ by
${\bm \sigma}({\bm{x}})$. The continuum limit $\epsilon \to 0$, $N\to
\infty$, and $n\to \infty$ with
$n\epsilon = \tau$ and $N\epsilon = T$ fixed, can then be taken. In doing so, we use
the formula
\begin{eqnarray}
\frac{1}{\epsilon}\,\frac{\partial}{\partial {\bm{v}}_n}\,\to\,
\frac{\delta}{\delta {\bm{v}}(\tau)}\,,
\end{eqnarray}
which transforms the partial derivative w.r.t. ${\bm{v}}_n$
into a functional derivatives w.r.t.  ${\bm{v}}(\tau)$. We thus
obtain the following functional differential evolution equation for
$W_\tau[{\bm{v}}]$ :
\begin{eqnarray}
\label{eq:W-velocity-equation}
\frac{\partial}{\partial \tau} W_\tau[{\bm{v}}]\,=\,
\frac{\delta}{\delta v^\alpha_\tau}\left[\frac{1}{2}
\frac{\delta}{\delta v^\beta_\tau}\Big(\chi^{\alpha\beta}({\bm{x}})
W_\tau[{\bm{v}}]\Big) - \sigma^\alpha({\bm{x}})W_\tau[{\bm{v}}]\right],
\end{eqnarray}
where ${\bm{v}}_t\equiv {\bm{v}}(\tau)$
and ${\bm{x}}\equiv {\bm{x}}[{\bm{v}}]$ given by Eq.~(\ref{xv0}).

Once the probability distribution over the paths $W_\tau[{\bm{v}}]$ is known,
it is of course easy to calculate probability
$P({\bm{x}},\tau)$ according to Eq.~(\ref{PW0}).
The only quantity which depends on $\tau$ on the r.h.s. of this equation
is
$W_\tau$. Taking the time derivative and using the equation
(\ref{eq:W-velocity-equation}), one easily reconstructs the Fokker-Planck
equation (\ref{eq:particle-FP-0}) satisfied by $P({\bm{x}},\tau)$.

One may also rewrite the probability 
distribution (\ref{eq:W-velocity-discrete0}) 
in the alternative discretization in which $\sigma$ and $\chi$ are
evaluated at ${\bm{x}}_{i}$, as in  Eq.~\eqref{eq:part-advanced}. As in
Sect. \ref{sec:same-phys-diff}, this requires using 
the relation (\ref{eq:A-sigchi}) between $\chi$ and ${\bm \sigma}$. 
One thus obtains:
\be\label{eq:W-velocity-discrete-adv}
W_n[{\bm{v}}]&=&\int \prod_{i=1}^N
\left(\frac{\epsilon}{2\pi}\right)^D d^D{\bm{p}}_i
\,{\rm exp}\bigg\{-{\rm i}\epsilon\sum_{i=1}^N  {\bm{p}}_i\cdot {\bm{v}}_i
\bigg\}\,\nn
&\times&\exp\bigg\{-\,\frac{\epsilon}{2}\sum_{i=1}^n 
\Big({\bm{p}}_i\cdot \chi({\bm{x}}_i)\cdot{\bm{p}}_i
+2{\rm i}\,{\bm{p}_i}\cdot \bm{\sigma}({\bm{x}}_i)\Big)\bigg\},\ee
which can be recast in continuum notations as follows:
\be\label{eq:W-velocity-adv}
W_\tau[{\bm{v}}]\,&=&\,\int\![{\rm
d}{\bm{p}}]\,
\exp\left\{ {-{\rm i}\int_{0}^\infty\!d\eta \,
{\bm{p}_\eta}\cdot {\bm{v}_\eta} } \right\}
\,\nn
&\times&\exp\left\{ -\frac{1}{2}\int_{0}^\tau\!d\eta
  \, \Big({\bm{p}}_\eta\cdot \chi({\bm{x}}_\eta)\cdot{\bm{p}}_\eta
+2{\rm i}\,{\bm{p}_\eta}\cdot \bm{\sigma}({\bm{x}}_\eta)\Big)\right\}
\ee
with ${\bm{x}}_\eta={\bm{x}}_\eta[{\bm{v}}]$ given by
the continuous form of Eq.~(\ref{eq:x-velocity}) :
\be\label{xv1}
{\bm{x}}_\eta[{\bm{v}}]={\bm{x}}_0+\int_0^\eta {\rm d}\tau' \,{\bm{v}}(\tau')
\ee
Although, for the reasons discussed in relation with
Eqs.~(\ref{eq:cont-A}) and (\ref{eq:P-prop-step-advanced}) above, such   
continuous notations can be ambiguous, they can be convenient. For
instance, when used with care, they allow simple formal operations, like
functional differentiations. In particular, it is easy to verify in this
way that Eq.~(\ref{eq:W-velocity-adv}) satisfies
equation (\ref{eq:W-velocity-equation}).
To this aim, it is convenient to use the alternative form of
Eq.~(\ref{eq:W-velocity-equation}) which is obtained after
commuting the two functional derivatives in its r.h.s.
through $\chi$ (i.e., the analog of Eq.~(\ref{eq:Ham2b})).
In this check,  the following identity is also useful:
\begin{equation}
\frac{\delta}{\delta v^\alpha_{\tau_1}}
\int_{0}^\tau\! d\eta\,F({\bm{x}}_\eta)\,=\,
\theta(\tau-\tau_1)\int_{\tau_1}^\tau\!d\eta\,
\partial^\alpha F({\bm{x}}_\eta),
\end{equation}
which vanishes for $\tau_1\geq \tau$.

\setcounter{equation}{0}
\section{Brownian motion in the space of Wilson lines}
\label{sec:BR-GROUP}

After all these preparations, we finally come to the study of the Brownian
motion in the space of Wilson lines, which is our main interest here.
The forthcoming discussion will in fact follow closely
that in Sect. \ref{sec:part-brownian}: as
we shall see, the evolution equations (\ref{eq:RGdef}) for $Z_\tau[U]$
and (\ref{RGEA}) for $W_\tau[\alpha]$, are the close
generalizations of Eqs.
(\ref{eq:particle-FP-0}) and (\ref{eq:W-velocity-equation}),
respectively.

The Wilson line $U({\bm{x}})$ is a field, with the coordinate ${\bm{x}}$
leaving in the transverse plane. In what follows, we shall often omit
these transverse coordinates in intermediate formulae, in order to
simplify the  notation and focus on
the true difficulties. These coordinates  will be reinserted in the final
formulae, or whenever ambiguities may arise from omitting them.
Similarly, we shall alternate between discrete and continuous time:  most
of the formulae are established for discrete times, and while many of
them have a straightforward continuous extension, one should keep in mind
that the use of continuous notations may hide ambiguities, as emphasized
in the previous section.

\subsection{Langevin equation and the $\alpha$-representation}

\label{sec:alpha}

The physical random variable in the evolution is the
elementary contribution $\alpha_\tau({\bm{x}})$ to
the classical field of the target arising from integrating out  gauge
field fluctuations in the the rapidity strip $[\tau,\tau+d\tau]$. Such a
contribution changes the  Wilson line according to ($\tau=n\epsilon,
\alpha_\tau {\rm d}\tau=\alpha_n\,\epsilon$):
\begin{equation}\label{LANGEVIN-gr}
U_i\,=\,U_{i-1}\,
{\rm e}^{-{\rm i}\epsilon\alpha_i^at^a}
,\qquad
U_i^\dagger\,=\,{\rm e}^{\mathrm{i}\epsilon\alpha_i^at^a}
\,U_{i-1}^\dagger .
\end{equation}
  A path in the space of $U$ fields is thus defined as the set of
values $\{U_{n}^\dagger,U_{n-1}^\dagger,\cdots,U_0^\dagger\}$, with :
\be\label{Wline-discrete}
U_{n}^\dagger\,=\,
{\rm e}^{{\rm i}\epsilon\alpha_{n}}U_{n-1}^\dagger
\,=\,\cdots\,=\,
{\rm e}^{{\rm i}\epsilon\alpha_{n}}{\rm e}^{{\rm
i}\epsilon\alpha_{n-1}}
\,\cdots\,{\rm e}^{{\rm i}\epsilon\alpha_{1}}\,U_0^\dagger
,\ee
In continuous notations, and with spatial coordinates included, a 
path is given by
Eq.~(\ref{U-path-alpha}) in Sect.~\ref{sec:restults-sketch}.
Note that in the previous formulae, and in most formulae
to come in this section, we set $g=1$ for simplicity.
To reintroduce the coupling constant $g$ in the
final results, it is enough to reinsert it in the 
exponent of the Wilson lines.

The random variables $\alpha_i^a({\bm{x}})$ at step $i$
are selected according to a
Gaussian probability distribution with the following characteristics:
\begin{equation}\label{fluct-ai}
\langle \alpha^a_{i}({\bm{x}})\rangle\,=\,
\sigma^a_{i-1}({\bm{x}})\,,\qquad
\langle\alpha^a_{i}({\bm{x}})\ \alpha^b_{i}({\bm{y}})\rangle\,=\,
\frac{1}{\epsilon}\,\chi^{ab}_{i-1}({\bm{x},\bm{y}})\,,
\end{equation}
These quantities are calculated in terms of the Wilson lines obtained at
the step
$i-1$, according to Eqs.~\eqref{eq:chidef} and~\eqref{eq:A-sigchi00}  
in Sect.~\ref{sec:RGE}. 

Equivalently, the variables $\alpha^a_{i}({\bm{x}})$ may be obtained by
solving the Langevin equation (with $e^{ab,l}_{i-1}$ given by
Eq.~(\ref{e}) with $U\equiv U_{i-1}$):
\begin{equation}\label{alpha-n}
\alpha^a_{i}({\bm{x}})\,=\,\sigma^a_{i-1}({\bm{x}})+
\,\int_{\bm{z}} e^{ab,l}_{i-1}({\bm{x},\bm{z}}) \
\nu^{b_,l}_{i}({\bm{z}})\,,
\end{equation}
where $\nu^a_{i}({\bm{x}})$ is an auxiliary Gaussian white noise:
\begin{equation}
\label{eq:noise-corr-gr}
\langle \nu^{a,l}_{i}({\bm{x}})\rangle\,=\,0,\qquad
\langle \nu^{a,l}_{i}({\bm{x}}) \nu^{b,k}_{j}({\bm{y}})
\rangle\,=\,\frac{1}{\epsilon}
\,\delta_{ij}\delta^{ab}\delta^{lk}\delta^{(2)}({\bm{x}-\bm{y}}),
\end{equation}
By averaging over the noise the solution of  Eq.~(\ref{alpha-n}) one
indeed recovers the expectation value  and correlator  of the random
variables
$\alpha_i^a({\bm{x}})$ as given by Eq.~(\ref{fluct-ai}).

The Langevin equation (\ref{alpha-n}) plays the same role for the
random walk on the group manifold as Eq.~(\ref{eq:Langevin-velocity})
for the random walk in a flat space, with the gauge field
$\alpha_i^a({\bm{x}})$ playing here the role of the velocity ${\bm{v}}_i$.
But in contrast to what happens in the simple example of
Sec.~\ref{sec:FP-velocity}, the relation between the ``coordinate''
$U_i$ and the ``velocity'' $\alpha_i$ is here non trivial, because of the
non-trivial geometry of the group manifold. However, as long as we
characterize the random walk by small displacements in the tangent
space, the special geometry of the group manifold plays essentially no
role (it only enters through the dependence of $\sigma$ and $\chi$ on the
Wilson lines).  The mathematical analogy with ordinary Brownian motion
in the velocity representation is then complete, and the formulae
obtained in Sect.~\ref{sec:FP-velocity} have an immediate translation
to the present problem.

Thus the weight function $W_n[\alpha]$ is given by the analog of
Eqs.~(\ref{eq:W-velocity-discrete0}) and (\ref{Pvfact}), namely:
\begin{equation}
\label{eq:W-alpha-discrete}
W_n[\alpha]=
\prod_{i=1}^n\,{\cal P}_i[\alpha]\prod_{j=n+1}^N \delta[\alpha_j]
\,,
\end{equation}
where
\begin{equation}\label{Pialpha1}
{\cal P}_i[\alpha]\equiv\left(\frac{\epsilon}{2\pi}\right)^{D/2}
\big({\rm det} \,\chi_{i-1}\big)^{-1/2}
\,{\rm e}^{- {\cal A}_i[\alpha]}
\,,
\end{equation}
\begin{equation}\label{Pialpha2}
{\cal A}_i[\alpha]=
\int_{\bm{x},\bm{y}}
\big(\alpha_{i}-\sigma_{i-1}\big)^a_{\bm{x}}
  [\chi^{-1}_{i-1}]^{ab}_{\bm{x y}} \big(\alpha_{i}-\sigma_{i-1}\big)^b_{\bm{y}},
\end{equation}
and
\be\delta[\alpha_i]\,\equiv\,\prod_{a}
\prod_{{\bm{x}}}\,\delta\big(\alpha^a_{i}({\bm{x}})\big).\ee
We assume here a suitable discretization of the transverse plane
with lattice points ${\bm{x}}$. In Eq.~(\ref{Pialpha1}), $D$ is the
total number of degrees of freedom, $D=(N_c^2-1)N_{\bm{x}}$ (i.e.,
the number of colors times the number of points in the transverse
lattice).
  Recall that $\sigma_{i-1}$, $\chi_{i-1}$ are functionals of $U_{i-1}$ so that
${\cal P}_i[\alpha]$
depend on all the variables $\alpha_1$, $\alpha_2$,
$\cdots \alpha_{i-1}$ (c.f. Eq.~(\ref{Wline-discrete})).
It is readily verified, by proceeding as indicated after 
Eq.~(\ref{Pvfact}), that the
weight function (\ref{eq:W-alpha-discrete}) is properly normalized:
\begin{equation}\label{measure}
\int [{\rm d}\alpha]\,\,W_n[\alpha]\,=\,1,\qquad
[{\rm d}\alpha]\,\equiv\,\prod_{i} \prod_{a}
\prod_{{\bm{x}}}\,{\rm d}\alpha^a_{i}({\bm{x}}).
\end{equation}

Since the inverse matrix $\chi$ is not immediately available, it is
convenient to introduce an extra integration over auxiliary momenta
$\pi_i^a({\bm{x}})$ and transform Eq.~(\ref{eq:W-alpha-discrete}) into
\begin{align}\label{Wn-alpha}
W_n[\alpha]=&\! \int [{\rm d}\pi]\,
\,{\rm exp}\Big\{\!\!-{\rm i}\epsilon\sum_{i=1}^N\int_{\bm{x}}
\pi^a_{i}\,\alpha^a_{i}
\Big\}
\nonumber \\
 &\times {\rm exp}\Big\{\!-\epsilon\sum_{i=1}^n
\left(\tfrac{1}{2}\int_{\bm{x},\bm{y}} \pi^a_{i}\, \chi_{i-1}^{ab}\, \pi^b_{i}
- {\rm i}\int_{\bm{x}}
\pi^a_{i} \sigma^a_{i-1}
\right)\Big\}.
\end{align}
where $[{\rm d}\pi] =\prod_{i=1}^N \prod_a\prod_{\bm{x}}\,(\epsilon/2\pi) {\rm d}\pi_i^a
({\bm{x}})$.  
It can then be easily verified by direct differentiation that $W_\tau[\alpha]$
is solution of the following functional evolution equation which is
the immediate translation of Eq.~(\ref{eq:W-velocity-equation}):
\begin{equation}\label{RGEA1}
\partial_\tau  W_\tau[\alpha] \,=\,\frac{\delta}{\delta \alpha^a_{\tau}({\bm{x}})}\left[
{1 \over 2} \,\frac{\delta}{\delta \alpha^b_{\tau}({\bm{y}})} \,
\Big(\chi^{ab}_{\bm{x y}}[\alpha] W_\tau[\alpha]\Big) \, - 
\sigma^a_{\bm{x}}[\alpha]W_\tau[\alpha]\right]\,.
\end{equation}
As expected, this is the same as the renormalization group equation in the
$\alpha$-representation, Eq.~(\ref{RGEA}).

\subsection{The Fokker-Planck equation in the space of $U$-fields}
\label{sec:U}

Let us now define $Z_n[U]$ as the probability density for the random
variable to take the value
$U$
at time $\tau= n\epsilon $, knowing that it is $U_{0}$ at time $0$.
This is the  analog of the  probability density
$P({\bm{x}},t)$ introduced in Sect. \ref{sec:part-brownian}.
  The normalization and initial conditions are written as
\be
\int  {\rm d}\mu(U)\,Z_\tau[U]\,=\,1\,\qquad
Z_0[U]\,=\,\delta_G(U,U_0)\,,\ee where ${\rm d}\mu(U)$ is the Haar measure
and $\delta_G(U,V)$ is the group-invariant $\delta$-function (see
Appendix).

According to the discussion in Sect. 2.3, the probability
$Z_n[U]$  can be calculated from the weight function
$W_n[\alpha]$, via Eq.~(\ref{eq:U-alpha-0})
(for more transparency, we suppress space coordinates) :
\be\label{ZvsW}
Z_n[U]=\int [{\rm d}\alpha]\,W_n[\alpha]\,
\delta_G\bigl(U^\dagger,\,U^\dagger_n[\alpha]\,\bigr),\ee
where $U^\dagger_n$ is given in terms of the $\alpha_i$'s by
Eq.~(\ref{Wline-discrete}).
(In fact, one could as
well replace $U^\dagger_n$ by $U^\dagger_N$ in the r.h.s. of this 
equation  since
$W_n[\alpha]$ ensures that $\alpha_j=0$ for any
$j>n$). In continuous notations, Eq.~(\ref{ZvsW}) becomes 
Eq.~(\ref{eq:U-alpha-0}).

By exploiting the factorized structure of $W_n[\alpha]$,
Eq.~(\ref{eq:W-alpha-discrete}), one can verify that $Z_n[U]$
satisfies a recurrence formula similar to Eq.~(\ref{P-exp}), that is,
\begin{equation}
\label{Z-recurr} Z_n[U]\,=\, \int \!\!  {\rm
  d}\mu(U_{n-1})\,\,Z_\epsilon[U|U_{n-1}]\,\,Z_{n-1}[U_{n-1}],
\end{equation}
and the elementary propagator $Z_\epsilon[U|U_{n-1}]$ is given by an equation
analogous to Eq.~(\ref{ZvsW}):
\begin{equation}\label{Zepsilon}
Z_\epsilon[U_i|U_{i-1}]\, = \,\int [d\alpha_i]\, {\cal P}_i[\alpha]\,\,\delta_G\Bigl(U^\dagger_i,\,
{\rm e}^{{\rm i}\epsilon\alpha_{i}}\,U^\dagger_{i-1}\Bigr).
\end{equation}
Note that for fixed $U_{i-1}$, the probability ${\cal P}_i[\alpha]$ is in
fact a function of $\alpha_i$ given by Eqs.~(\ref{Pialpha1}) and
(\ref{Pialpha2}).  We expect $Z_\epsilon[U_i|U_{i-1}]$ to differ from unity
by terms which vanishes as $\epsilon\to 0$. In fact the deviation is of order
$\epsilon$, as we now show. To proceed, we note that the weight function
${\cal P}_i[\alpha]$ is a Gaussian of width $\epsilon$ in the variable $\epsilon\alpha$, and
we perform an expansion of the group $\delta$-function in
Eq.~(\ref{Zepsilon}) up to quadratic order in $\epsilon\alpha_i\sim \sqrt{\epsilon}$.  To perform this expansion, we
apply the identity (\ref{TaylorU}) derived in Appendix and obtain, in
direct generalization of Eq.~\eqref{eq:delta_expansion},
\begin{align}
\label{delta-exp}
\delta_G\Bigl(U^\dagger_i,\, &
{\rm e}^{{\rm i}\epsilon\alpha_{i}}\, U^\dagger_{i-1}\Bigr)
 =  \delta_G\Bigl({\rm e}^{-{\rm i}\epsilon\alpha_{i}}\,U^\dagger_i,\,
U^\dagger_{i-1}\Bigr)\,=\,{\rm e}^{- \epsilon\alpha_{i}^a\nabla^a}
\delta_G\Bigl(U^\dagger_i,\,U^\dagger_{i-1}\Bigr)
\nonumber \\
 \approx &\Bigl(1-\epsilon\alpha_{i}^a\nabla^a+\,\frac{\epsilon^2}{2}
\alpha_{i}^a\alpha_{i}^b\nabla^a \nabla^b
\Bigr)\delta_G\Bigl(U^\dagger_i,\,U^\dagger_{i-1}\Bigr)\,,
\end{align}
where the Lie derivatives act on $U_i$ (not on $U_{i-1}$).  By
inserting this expression in Eq.~(\ref{Zepsilon}) and performing the
integral over $\alpha_i$, we finally arrive at
\begin{equation}\label{Zepsilon-exp}
Z_\epsilon[U|U_{n-1}]\,=\,\Bigl\{1+\epsilon\Bigl(-
\sigma^a_{n-1}\nabla^a+\,
\frac{1}{2}\chi^{ab}_{n-1}\nabla^a \nabla^b\Bigr)
\Bigr\}\delta_G\Bigl(U^\dagger,\,U^\dagger_{n-1}\Bigr)  +{\cal
O}(\epsilon^{3/2}).
\end{equation}
The steps needed to obtain the equation satisfied by $Z_\tau[U]$ in the
continuum limit are now identical to those leading to the
Fokker-Planck equation (\ref{eq:particle-FP-0}). Reinstating
coordinate dependence, one gets
\begin{equation}
\label{RGE-Z}
\partial_\tau Z_\tau[U]\,=\,
{1 \over 2}\, \nabla^a_{\bm{x}}\nabla^b_{\bm{y}}\,
\chi^{ab}_{\bm{x y}}[U] Z_\tau[U]\, - \,
\nabla^a_{\bm{x}}\,
\sigma^a_{\bm{x}}[U] Z_\tau[U]\,,
\end{equation}
which is Eq.~(\ref{eq:RGdef}).

On a more formal level, this equation could have been simply obtained
by taking the time derivative of Eq.~(\ref{eq:U-alpha-0}) relating
$Z_\tau[U]$ and $W_\tau[\alpha]$.  Noting that only $W_\tau$ depends on time in the
r.h.s. of this equation, and using Eq.~(\ref{RGEA1}) for $\partial_\tau W_\tau$,
one obtains:
\begin{align}
\label{RGE-ZW}
\partial_\tau & Z_\tau[U] =
  \int [{\rm d}\alpha]\,\partial_\tau W_\tau \,
\delta_G\Bigl(U^\dagger,{\rm T\,e}^{{\rm i}\int\limits_{0}^\infty  
d\eta\, \alpha_{\eta}} \,U_0^\dagger\Bigr)
\nonumber \\
=&\int [{\rm d}\alpha]\, W_\tau[\alpha]\,\left\{
\frac{1}{2} \,\chi^{ab}_{\bm{x y}}[\alpha]\frac{\delta^2}{\delta
\alpha^a_{\tau}({\bm{x}}) \delta \alpha^a_{\tau}({\bm{y}})}\,+\,\sigma^a_{\bm{x}}[\alpha]
\frac{\delta}{\delta \alpha^a_{\tau}({\bm{x}})}\right\}
\delta_G\Bigl(U^\dagger,{\rm T\,e}^{{\rm i}\int\limits_{0}^\tau  d\eta\, 
\alpha_{\eta}} \,U_0^\dagger\Bigr)
\end{align}
In the second line, we have performed integrations by parts w.r.t.
$\alpha$, and then restricted the integral in the exponent of the Wilson
line to $\eta\leq\tau$, taking advantage that the weight function $W_\tau[\alpha]$
imposes $\alpha^a_\eta=0$ for any $\eta>\tau$ (c.f. Eq.~(\ref{eq:W-alpha-discrete})).
We then observe that the functional derivatives acting on this
particular Wilson line have the same effect as the Lie derivative
$\nabla^a$. Setting $V^\dagger\equiv {\rm T} \exp({\rm i}\int_{0}^\tau\! d\eta\,\alpha_\eta )U_0^\dagger$, we have
indeed (see Eq.~(\ref{lieUU})):
\begin{equation}\label{DIFFU} \frac{\delta}{ \delta
\alpha^a_{\tau}({\bm{y}})} V^\dagger({\bm{x}})= {\rm
   i}t^a\,V^\dagger({\bm{y}})\,  \delta^{(2)}({\bm{x}-\bm{y}})= 
\nabla^a_{\bm{y}}V^\dagger({\bm{x}})\,.
\end{equation}
This allows us to trade the functional derivatives in
Eq.~(\ref{RGE-ZW}) for Lie derivatives, e.g.,
\begin{equation}
{\delta \over {\delta \alpha_\tau^a}}
\delta_G\Bigl(U^\dagger,{\rm T\,e}^{{\rm i}\int\limits_{0}^\tau \! d\eta\, \alpha_\eta }U_0^\dagger\Bigr)=
\nabla^a_V
\delta_G\bigl(U^\dagger,V^\dagger\bigr)=-\nabla^a_U
\Bigl(U^\dagger,{\rm T\,e}^{{\rm i}\int\limits_{0}^\tau \! d\eta\, \alpha_\eta }U_0^\dagger\Bigr),\,\,
\end{equation}
and thus to immediately recover Eq.~(\ref{RGE-Z}).

\subsection{Path integrals in the space of $U$-fields}

We now return to Eq.~(\ref{ZvsW}) and use the known expression
(\ref{Wn-alpha}) for $W_n[\alpha]$ to derive a path-integral representation
for $Z_n[U]$ directly in terms of the $U_i$'s, rather than the
$\alpha_i$'s. 

We proceed as usual, divide the rapidity interval $\tau$
into
$n$ small elements of length $\epsilon$, and use repeatedly
Eq.~(\ref{Z-recurr}) to deduce that
\begin{equation}\label{Z-recurr1}
Z_\tau[U] \,=\,\int \prod_{i=1}^{n-1} [{\rm d}\mu(U_i)]\,\,
Z_\epsilon[U|U_{n-1}]\,\,Z_\epsilon[U_{n-1}|U_{n-2}]\,\cdots\,
Z_\epsilon[U_1|U_0] .
\end{equation}
Next, we define
\begin{equation} \label{theta}
U^\dagger_iU_{i-1}\,\equiv\,{\rm e}^{{\rm
i}\epsilon\omega_i^at^a},\qquad {\rm
i}\epsilon\omega_i^a\,=\,
2\,{\rm tr}\,\,t^a\ln\bigl(
U^\dagger_iU_{i-1}\bigr).
\end{equation}
The $\omega_i$'s are uniquely determined as long as the matrix
$U^\dagger_iU_{i-1}$ is sufficiently close to the unit matrix, which will be
the case.

Consider then the elementary propagator $Z_\epsilon[U_i|U_{i-1}]$ as given by
Eq.~(\ref{Zepsilon}). Using the parametrization (\ref{theta}), one can
express the group $\delta$-function in Eq.~(\ref{Zepsilon}) as an ordinary
$\delta$-function for the random variable $\alpha^a_i$ over which we need to
integrate (see Eq.~(\ref{deltaf}) in Appendix):
\begin{equation}\label{deltaG}
\mu(\epsilon\omega_i)\,\delta_G\bigl({\rm e}^{{\rm
i}\epsilon\omega^a_it^a}\!,\, {\rm
e}^{i\epsilon\alpha_{i}^ct^c}\bigr)\,=\,
\delta(\epsilon\omega_i-\epsilon\alpha_{i}).
\end{equation}
It is then straightforward to perform
the integral over $\alpha_{i}^a$ in Eq.~(\ref{Zepsilon}), and obtain:
\begin{align}
\label{elemZ}
Z_\epsilon[U_i|U_{i-1}] = &\ \frac
{\big({\rm det} \,\chi_{i-1}\big)^{-1/2}}
{(2\pi \epsilon)^{D/2}}\,\,\,\prod_{\bm{x}}
[\mu(\epsilon\omega_i({\bm{x}}))]^{-1}
\nonumber \\
  & \times \ \exp\Big\{-\frac{\epsilon}{2}\,\int_{\bm{x},\bm{y}}
\big(\omega_i-\sigma_{i-1}\big)^a_{\bm{x}}
  \big[\chi_{i-1}^{-1}\big]^{ab}_{\bm{x y}} 
  \big(\omega_i-\sigma_{i-1}\big)^b_{\bm{y}}
\Big\},
\end{align}
where it is understood that $\omega^a_i({\bm{x}})$ is a function of
$U_i({\bm{x}})$ and $U_{i-1}({\bm{x}})$ obtained by inverting the
relations (\ref{theta}).  Eq.~(\ref{elemZ}) is the analog of
Eq.~(\ref{eq:P-prop-epsilon}), but it is actually more complicated than the
latter because $\omega_i$ is  non-linearly related to the
neighboring group elements $U_i$ and $U_{i-1}$ (cf. Eq.~(\ref{theta})),
as required by the geometry of the group manifold.

Eq.~(\ref{elemZ}) implies that, typically,
$\epsilon\omega^a\sim \sqrt{\epsilon}$, which is consistent with
Eq.~(\ref{deltaG}) and the  previous estimates for $\alpha_{i}$
(cf. Eq.~(\ref{fluct-ai})). This shows that, in order to control
Eq.~(\ref{elemZ}) to the accuracy of interest (i.e., up to and
including terms of order $\epsilon$),
it is enough to evaluate the measure factor 
$\mu(\epsilon\omega_i)$ to quadratic order in 
$\epsilon\omega_i$. An explicit calculation to be presented in
Appendix \ref{sec:App-Lie} yields  (cf. Eq.~(\ref{mu-2nd})):
\begin{equation} \label{mu}
[\mu(\epsilon\omega_i)]^{-1}\,=\,1+\,
\epsilon^2\,\frac{N_c}{24}\omega^a_i\omega^a_i+
{\cal O}(\epsilon^{3/2})\,=\,{\rm
e}^{\epsilon^2\frac{N_c}{24}\omega^a_i\omega^a_i}+ {\cal
O}(\epsilon^{3/2}).
\end{equation}

We are now in a position to write $Z_\tau[U]$ as a path integral:
\be\label{ZtaudeU}
Z_\tau[U] \,=\,\int_{U_0}^U  [{\rm d}\mu (U)]\,\, \exp\{-{\cal A}[U]\},
\ee
where
\be\label{intsurU}
  [{\rm d}\mu( U)]\equiv
 \prod_{i=1}^{n-1}{\big({\rm det} \,\chi_{i-1}\big)^{-1/2}}
{(2\pi \epsilon)^{-D/2}}\,\,\,[\mu(\epsilon\omega_i)]^{-1}\, {\rm d}\mu (U_i)
\ee
and
\begin{equation}
{\cal A}[U]=-\frac{\epsilon}{2}\,\sum_{i=1}^n\,\int_{\bm{x},\bm{y}}
\big(\omega_i-\sigma_{i-1}\big)^a_{\bm{x}}
  \big[\chi_{i-1}^{-1}\big]^{ab}_{\bm{x y}} 
  \big(\omega_i-\sigma_{i-1}\big)^b_{\bm{y}}
.
\end{equation}

Again, one can trade the
 matrix $\chi^{-1}$ for $\chi$ itself by introducing an
additional integration over auxiliary momentum variables:
\begin{equation}\label{path-Z}
Z_\tau[U]\,=\,
\int \prod_{i=1}^{n-1}
   d\mu (U_i) \int \prod_{i=1}^{n}d\pi_i\,\,{\rm
e}^{-\epsilon\sum_{i=1}^{n} {\cal L}_i[ \pi,U]},
\end{equation}
with
\begin{equation}
\label{LU}
{\cal L}_i[ \pi,U]\,\equiv\,-\,
\frac{\epsilon N_c}{24}\!\int_{\bm{x}}\omega_{i}^a\omega_{i}^a
\,+\,{\rm i}\int_{\bm{x}}
\pi_{i}^a\bigl(\omega_{i}-\sigma_{i-1}\bigr)^a\,
+\,\frac{1}{2}\int_{\bm{x},\bm{y}}\!
\pi_{i}^a\, \chi^{ab}_{i-1} \,\pi_{i}^b\,,
\end{equation}
and we have also used Eq.~\eqref{mu}.
To reintroduce the coupling constant $g$ in the previous
formulae, it is enough to insert
a factor $1/g$ in the r.h.s. of Eq.~\eqref{theta} for $\omega_{i}^a$.

The role of the measure term (i.e., the first term in the r.h.s. of
Eq.~(\ref{LU})) should be quite clear with the following remark:
Assume that we decide to evaluate the path-integral (\ref{ZtaudeU}) by
using the ``relative angles'' $\omega_i$ (in the sense of
Eq.~(\ref{theta})) as the integration variables. Then, as obvious from
Eq.~(\ref{intsurU}), the measure terms simply drop out:
$[\mu(\omega_i)]^{-1}\, {\rm d}\mu (U_i)={\rm d}\omega_i$, and Eq.~(\ref{ZtaudeU})
reduces to Eq.~(\ref{ZvsW}) with $W_n[\alpha]$ of
Eq.~(\ref{eq:W-alpha-discrete}).

\setcounter{equation}{0}
\section{Conclusions}
\label{sec:conclusions}

We have established the relationships between the two forms of
the renormalization group equation towards small $x$ 
which are available in the literature. These have been 
obtained in different approaches, and, in spite of some formal
similitudes, they rely on different mathematical structures:
the $U$- and the $\alpha$-representations of this paper. 
What we have shown here is that these two representations
are equivalent and describe the same basic statistical
process: a random walk in the space of Wilson lines,
with $\ln 1/x$ playing the role of ``time'', 
$\alpha$ that of the ``velocity'', and $U$ being
the ``position'' on the group manifold.

The Langevin description of this random walk provides a natural bridge
between the two representations: At step $i$, the gauge
field $\alpha_i$ is randomly chosen according to the gaussian probability
distribution (\ref{fluct-ai}),  then used to update the SU$(N)$ group
element $U_i$, cf. eq.~(\ref{LANGEVIN-gr}), which is in turn used to
compute the probability law for selecting the ``velocity'' $\alpha_{i+1}$
at step $i+1$, and so on. This Langevin process provides an algorithm for
the quantum evolution and stays at the basis of a numerical study of the
renormalization group  equations which is currently in progress
\cite{HWR}. Note that the results of such
Langevin simulations will not only give access to the $x$ dependence of
cross sections like the $\gamma^*A$ cross sections of
Eq.~\eqref{eq:sigma_tot}, they also provide the basis for an extension to
the determination of initial conditions in $A A$ collisions with a
calculated $x$ dependence along the lines of~\cite{Kovner:1995ja,KV00}.

In addition to its suitability for practical implementations, the
Langevin formulation brings also important conceptual clarifications,
as it most prominently exhibits the statistical nature of the underlying
physics: The renormalization group at small
$x$ is driven by successively integrating out gluonic fluctuations within a
small slice in $\ln 1/x$ around a static (and typically large) background field,
which represents the quantum modes integrated in previous steps,
and which are effectively ``frozen''. 
The ensuing coarse graining in the longitudinal
direction is accompanied by a loss of quantum coherence, allowing
eventually a description in terms of a classical stochastic process.

The particular structure of the stochastic equations which emerge from
the renormalization group has allowed us to avoid discretization
ambiguities which generally arise for random walks in inhomogeneous
media. We have utilized the so-called Ito discretization,
which allows one to construct the new
variables in a ``causal'' manner from the knowledge of the old ones, and
we have verified that, with this choice, the  Langevin description
reproduces indeed the expected renormalization group equations --- which
here emerge as {\it Fokker-Planck} equations --- in the continuum limit.
Once things are properly formulated within one  discretization,
other prescriptions can be used as well, and we have briefly
elaborated on some alternatives.

The discretization that we have used has also the virtue to 
greatly facilitate the construction of path-integral representations
for the solutions to the Fokker-Planck equations. Such path-integrals can 
be the starting point for alternative
numerical implementations. It would be interesting in this context to
explore the possibility to rewrite this  path-integral in a form which is
local in the transverse coordinates, possibly after introducing
auxiliary fields, as in Ref. \cite{Balitsky2001}.
But path-integrals can be also useful for  analytical studies.
and could give us more insight into the  qualitative
features of the non-linear physics at small $x$.
Although to some extent formal, the explicit path-integral solutions
that we have obtained have the merit to exhibit important structural
properties, like non-locality (in both longitudinal and 
transverse coordinates), non-linearity, and colour structure.
Less formal analytic results could be obtained
by further evaluating the path integrals, e.g., via
saddle-point or mean-field approximations.
In particular, the self-consistent mean-field approximations
discussed in Ref. \cite{SAT} can be easily reformulated within the
present approach.

\section*{Acknowledgments}

Two of us (J.-P. Blaizot and E. Iancu) would like to thank the KITP in
the University of California at Santa Barbara, and the organizers of
the program {\it ``QCD and Gauge Theory Dynamics in the RHIC Era''},
for hospitality at various stages during the completion of this work.
We would like to acknowledge useful conversations with Ian Balitsky,
Larry McLerran, Al Mueller, Robi Peschanski and A. Sch{\"a}fer.  This
research was supported in part by the National Science Foundation
under Grant No. PHY99-07949. HW was supported by a DFG Habilitanden
fellowship.

\setcounter{equation}{0}
\appendix
\section{Calculations on Lie group: useful formulae}
\label{sec:App-Lie}

We collect in this appendix a few  identities which are useful in calculations on
group manifolds (see, e.g., Refs. \cite{Miller,Zinn99} for more details).

We consider SU(N) matrices $U$. These depend on $N^2-1$ real parameters
$\omega^a$, and can be written in the exponential form as:
\be
U(\omega)={\rm e}^{-{\rm i}\omega^a t^a}\equiv {\rm e}^{-{\rm
i}\,\bfomega}\qquad U^\dagger (\omega) ={\rm e}^{{\rm
i}\,\bfomega}
\ee
where the $t^a$'s are the infinitesimal generators
normalized as Tr$(t^at^b)=d_R\delta^{ab}$, with 
$d_R$ the dimension of the representation considered. When needed, we shall distinguish
the adjoint representation by a
tilde:
$\tilde U(\omega)=
 {\rm e}^{-{\rm i}\tilde\omega}$.

Consider first the case where these parameters depend on a single variable, the
``time'' 
$\tau$. When $\tau$ varies, the matrix $U_\tau$
describes a curve on the group manifold.  Consider a variation of the matrix $U$
when 
$\tau$ increases by a small amount 
${\rm d}\tau$.  We can write:
$
U_{\tau+{\rm d}\tau}=U_\tau+{\rm d}U_\tau$, and note that
$U^\dagger_\tau {\rm d}U_\tau=U^\dagger_\tau U_{\tau+{\rm d}\tau}-1$ is an element of
the Lie algebra that we can write as
$
U^\dagger_\tau {\rm d}U_\tau=-iT^a\alpha^a {\rm d}\tau
$
where $\alpha^a(\tau)$ are the components of a vector tangent to the curve. Clearly
${\rm d}U_\tau=-iU_\tau T^a\alpha^a {\rm d}\tau $, or equivalently 
\be
{\rm
d}U_\tau^\dagger=i\,\alpha^a\,T^a  \, U_\tau^\dagger\, {\rm d}\tau
\ee which allows us to write
$U_\tau^\dagger$ as a time ordered exponential:
\be\label{Lie0}
U_\tau^\dagger={\rm T} \exp\{i\int_0^\tau {\rm d}\eta\, \alpha^a(\eta)
\,T^a\}\,U_0^\dagger
\ee
Note that it is the matrix $U^\dagger$, rather than $U$ itself, which enters our
discussion in the main text. Note also that $\alpha^a\ne {\rm d}\omega^a/{\rm d}\tau$.

Consider next an arbitrary function of $U_\tau^\dagger$. We may write:
\be
\frac{{\rm d}f(U^\dagger_\tau)}{{\rm d}\tau}\,=\,\big(i \alpha^aT^aU^\dagger\big)_{i
j}\,
\frac{\delta}{\delta U^\dagger_{i j}}\,f[U^\dagger]\equiv\alpha^a\, \nabla^a
f(U^\dagger),
\ee
and therefore
\be\label{Lief}
f(U^\dagger)={\rm T}\exp\left\{  \int_0^\tau d\eta\, \alpha^a(\eta)
\,\nabla^a\}\right\} f(U_0^\dagger)
\ee
The operators $\nabla^a$ introduced above, commonly referred to as
Lie derivatives, satisfy the following
relations:
\be\label{lieUU}
 \nabla^a
U= -{\rm i}U\,T^a \,,\quad
\nabla^a
U^\dagger= {\rm i}T^a\,U^\dagger\,\,,
\quad 
[\nabla^a ,\,\nabla^b] \,=\, f^{a b c}\,
\nabla^c\,,\ee
and  form therefore a representation of the Lie algebra.

If one chooses  $\alpha$ to be constant in eqs.~(\ref{Lie0}) and (\ref{Lief}), one
gets the following useful formula:
\be\label{TaylorU}
f({\rm e}^{{\rm i}\alpha^aT^a}U_0^\dagger)={\rm e}^{\alpha^a\nabla^a}f(U_0^\dagger)
\ee

On the group manifold, one may define a metric by
$
{\rm d} s^2={\rm Tr} \left( {\rm d}U {\rm d}U^\dagger \right)
$
or, in terms of the group parameters $\omega^a$:
\be
{\rm d} s^2=g_{ab}\,\omega^a\omega^b= {\rm Tr} \left( \frac{{\rm d}U}{{\rm
d}\omega^a}\frac{ {\rm d}U^\dagger}{{\rm d}\omega^b}
\right)=\left(LL^t\right)_{ab}\,,
\ee
where we have set
\be
 U^\dagger
\frac{\partial}{\partial\omega^a}\,U\,=\,-{\rm i}L_{ab}(\omega)\,T^b\,,
\ee
and $L^t_{ab}=L_{ba}$.
From this one gets easily the Haar invariant measure
\be
{\rm d}\mu(U)=\mu(\omega)\Pi_a {\rm d}\omega_a\qquad\mu(\omega)= \sqrt{|{\rm det}
\,g(\omega)|}\,.
\ee
To calculate $L_{ab}$ we use:
\be\label{omega-deriv}
U^\dagger\,
\frac{\partial U}{\partial\omega^a}=\int_0^1 ds\,
{\rm e}^{{\rm i}s \bfomega}\big(-{\rm i}T^a\big){\rm e}^{-{\rm i}s\bfomega}
\,=\,-{\rm i}\int_0^1 ds\,\left( {\rm e}^{-{\rm i}s {\bf \tilde \omega}}\right)_{ab}
\, T^b.\ee Note the emergence
of adjoint-representation matrices ${\bf \tilde\omega}$ in the r.h.s.'s of the above
formulae. Since matrices
in the adjoint representation are real,  $
({\rm e}^{-{\rm i}s{\bf \tilde\omega}})_{ab} =
({\rm e}^{{\rm i}s{\bf \tilde\omega}})_{ba}$). One can then write
\be
\big(L(\omega) L^t(\omega)\big)_{ab}
=\int_0^1 ds\int_0^1 d\lambda\,
\big({\rm e}^{{\rm i}(s-\lambda)\tilde\omega}\big)_{ab}=
\biggl(\frac{1}{2}\int_{-1}^1ds\,{\rm e}^{\frac{{\rm i}}{2}s\tilde\omega}
\biggr)^2_{ab}, \ee
and obtain $\mu(\omega)=|\det({\cal J}_{ab}(\omega))|$ with 
\be\label{mu-omega}
{\cal J}_{ab}(\omega)\,\equiv\,\frac{1}{2}\int_{-1}^1ds\,
\big({\rm e}^{\frac{{\rm i}}{2}s\tilde\omega}\big)_{ab}\,=\,\sum_{n\ge 0}\frac{
(-)^n\big(\tilde\omega^{2n}\big)_{ab}}{2^{2n}(2n+1)!}\,=\,\biggl(\frac{\sin(\tilde\omega/2)}
{\tilde\omega/2}\biggr)_{ab}\,.\ee

This expression allows in particular the computation of  $\mu(\omega)$ to
quadratic order in $\omega$, as required in Sect.~4.3:
\be\label{mu-2nd}
\mu(\omega)\,=\,\det {\cal J}(\omega)\,=\,{\rm e}^{{\rm Tr}\ln
{\cal J}(\omega)}\,=\, {\rm e}^{-\frac{N}{24}\omega^a\omega^a}+
{\cal O}(\omega^{3}).\ee

Integrals over the group manifold are  
typically of the form:
\be
\int {\rm d}\mu(U) \, f(U)=\int \Pi_a d\omega_a\, \mu(\omega) f({\rm
e}^{-i\,\bfomega}).
\ee
A particularly useful tool is the group $\delta$-function defined by
\begin{equation}
  \label{eq:group-delta}
  \int {\rm d}\mu (U) f[U] \,\delta_G[U,V] = f[V]
\end{equation}
with obvious invariance properties, like $\delta_G[U,V] = \delta_G[U V^\dagger,1] =
\delta_G[1,U^\dagger V]$, which follow directly from the invariance of the Haar
measure ${\rm d}\mu (U)$. It will useful to relate this group $\delta$-function to the
ordinary $\delta$-function. To do so, consider the integral:
\be
f({\rm e}^{-i\bftheta})&=&\int {\rm d}\mu(U) \, f(U)\,\delta_G(U,{\rm
e}^{-i\bftheta})\nn &=&\int
\Pi_a d\omega_a\, \mu(\omega) \,f({\rm e}^{-i\bfomega})\,\delta_G({\rm
e}^{-i\bfomega},{\rm e}^{-i\bftheta})
\ee
On the other hand, we have
\be
f({\rm e}^{-i\bftheta})=\int \Pi_a
{\rm d}\omega_a\,f({\rm e}^{-i\bfomega})\,\delta(\omega-\theta),
\ee
with $\delta(\omega-\theta)\equiv \Pi_a \delta(\omega^a-\theta^a)$.
Combining these two equations, we obtain the following identity:
\be\label{deltaf}
\mu(\omega)\delta_G({\rm e}^{-i\bftheta},{\rm e}^{-i\bfomega})=\delta(\omega-\theta)
\ee
Further useful formulae involve Lie derivatives. First we may use 
 e.q.~(\ref{omega-deriv}) together with the properties (\ref{lieUU}) of Lie derivativs
to obtain:
\be
\frac{\del U}{\del \omega^a}=L_{ab}\nabla^b\,U\qquad
\frac{\del U^\dagger}{\del \omega^a}=L_{ab}\nabla^b\,U^\dagger.
\ee
From these relations, and again e.q.~(\ref{omega-deriv}),  one easily deduces the
following useful identities:
\be
\left.\nabla^a
f(U)\right|_{U=1}=\frac{\del}{\del\omega^a}\left.f(U(\omega))\right|_{\omega=0}\,,
\ee
\be
\frac{1}{2}
\left.\left(\nabla^a\nabla^b+\nabla^b\nabla^a\right)f(U)\right|_{U=1}=\frac{\del^2}
{\del\omega^a\del\omega^b}\left.f(U(\omega))\right|_{\omega=0}\,.
\ee
By using these fromulae and the relation obtained above for the group
$\delta$-function, one can  also derive the following identities:
\be
\mu(\omega)\nabla^a
\delta_G(U(\omega),1)=\frac{\del}{\del\omega^a}\delta(\omega)\,,
\ee
\be
\mu(\omega)
\chi^{ab}\nabla^a\nabla^b\delta_G(U(\omega),1)=\chi^{ab}\frac{\del^2}
{\del\omega^a\del\omega^b}\delta(\omega)\,,
\ee
for any symmetric $\chi^{ab}$.

All the previous formulae are easily extended to the case where the matrices $U$ are
fields $U({\bf x})$ in the transverse coordinates ${\bf x}$. Thus for instance, Lie
derivatives become
\be
 \nabla_{\bm{x}}^af[U]\,=\,\big(-iU({\bm{x}})t^a\big)_{i j}\,
\frac{\delta}{\delta U_{i j}({\bm{x}})}\,f[U]\,.\ee
We have also:
\be\label{lieUUx}
 \nabla_{\bm{x}}^a
U_{\bm{y}}= -{\rm i}U_{\bm{y}}\,t^a \,\delta_{\bm{x y}}\,,\quad
\nabla_{\bm{x}}^a
U^\dagger_{\bm{y}}= {\rm i}t^a\,U^\dagger_{\bm{y}}\, \delta_{\bm{x y}}\,,
\quad 
[\nabla_{\bm{x}}^a ,\,\nabla_{\bm{y}}^b] \,=\, f^{a b c}\,
\nabla_{\bm{y}}^c\delta_{\bm{x y}}\,,\ee
where
 $\delta_{\bm{x y}}\equiv \delta^{(2)}({\bf x}-{\bf y})$.

\include{bibliography}

\end{document}